\documentclass[11pt]{article}

\usepackage{fullpage}
\usepackage{graphicx}
\usepackage{amsmath}
\usepackage{amssymb}
\usepackage{hyperref}
\usepackage{fix-cm}

\renewenvironment{thebibliography}[1]{%
\begin{oldthebibliography}{#1}%
\setlength{\parskip}{0ex}%
\setlength{\itemsep}{0ex}%
}%
{%
\end{oldthebibliography}%
}

\newcommand{\str}{\rule{0ex}{2.7ex}}

\begin{document}
~\vspace{0.3cm}

\centerline{\LARGE Semiclassical approach to discrete symmetries in quantum chaos}
\vspace{0.8cm}
\centerline{\large Chris Joyner\footnote{chris.joyner@bristol.ac.uk}, Sebastian M\"uller and Martin Sieber}
\vspace{0.3cm}
\centerline{School of Mathematics, University of Bristol, Bristol BS8\,1TW, UK}

\vspace{1.5cm}
\centerline{\bf Abstract}
\vspace{0.4cm}
\noindent
We use semiclassical methods to evaluate the spectral two-point correlation function of quantum chaotic systems with discrete geometrical symmetries. The energy spectra of these systems can be divided into subspectra that are associated to irreducible representations of the corresponding symmetry group. We show that for (spinless) time reversal invariant systems the statistics inside these subspectra depend on the type of irreducible representation. For real representations the spectral statistics agree with those of the Gaussian Orthogonal Ensemble (GOE) of Random Matrix Theory (RMT), whereas complex representations correspond to the Gaussian Unitary Ensemble (GUE). For systems without time reversal invariance all subspectra show GUE statistics. There are no correlations between non-degenerate subspectra. Our techniques generalize recent developments in the semiclassical approach to quantum chaos allowing one to obtain full agreement with the two-point correlation function predicted by RMT, including oscillatory contributions.
\vspace{1.5cm}

\section{Introduction}

According to the random matrix conjecture \cite{BGS84} the energy spectra of quantum chaotic systems have universal statistical properties that depend only on the symmetries of the system and agree with predictions from random matrix theory (RMT). Usually one considers systems without geometrical symmetries. In this case one obtains agreement with the Gaussian Unitary Ensemble (GUE) if time reversal invariance is broken. The statistics of time reversal invariant systems agrees with the Gaussian Orthogonal Ensemble (GOE) if the time reversal operator squares to one, and with the Gaussian Symplectic Ensemble (GSE) if it squares to minus one; the latter situation can occur in spin systems.

In this paper we want to consider individual chaotic systems with discrete geometrical symmetries. The energy levels of these systems fall into subspectra associated to irreducible representations of the underlying symmetry group. For example, in systems with a reflection symmetry
the subspectra correspond to eigenfunctions that are either even or odd under reflection. Moreover,
if the system is chaotic each subspectrum individually obeys RMT statistics. One remarkable property, first observed by Leyvraz, Schmit and Seligman \cite{Ley96}, of symmetric systems is the ability to generate GUE statistics within certain subspectra regardless of the time-reversal properties of the full system.
A partial explanation of these observations was given by Keating and Robbins \cite{Rob89,Kea97}
who used semiclassical analysis to
evaluate (via its Fourier transform) the two-point correlation functions
\begin{equation}
R(\epsilon)
= \frac{1}{{\bar\rho}^2} \left< \rho\left(E + \frac{\epsilon}{2\pi \bar{\rho}} \right) \rho\left(E - \frac{\epsilon}{2\pi \bar{\rho}} \right) \right> -1\label{Correlation_Function}
\end{equation}
associated to each subspectra.
Here $\rho(E)=\sum_n\delta(E-E_n)$ is the corresponding level density, $\bar\rho$ is the mean level density, and $\left\langle \cdot \right\rangle$ denotes an energy average.
The GUE and GOE predictions for $R(\epsilon)$ are \cite{Meh04,Haa10}
\begin{eqnarray}
R_{\rm GUE}(\epsilon)&=&-\left( \frac{\sin \epsilon}{\epsilon} \right)^2={\rm Re}\left(-\frac{1}{2\epsilon^2} + \frac{e^{2i\epsilon}}{2\epsilon^2}\right)\label{GUE_Expansion}\\
R_{\rm GOE}(\epsilon)&=&-\left( \frac{\sin \epsilon}{\epsilon} \right)^2 +
\left( \int_{0}^{\epsilon} \frac{\sin y}{y} d y - \frac{\pi}{2} \mathrm{sgn}(\epsilon) \right) \,
\left(\frac{\cos \epsilon}{\epsilon} - \frac{\sin \epsilon}{\epsilon^2} \right)\nonumber\\
&\sim&{\rm Re}\left(-\frac{1}{\epsilon^2} + \sum_{k=3}^{\infty}\frac{(k-3)!(k-1)}{2i^k}\frac{1}{\epsilon^k}
+\sum_{k=3}^{\infty}\frac{(k-3)!(k-3)}{2i^k}\frac{e^{2i\epsilon}}{\epsilon^k}\right)
\label{GOE_Expansion}\;.
\end{eqnarray}
Here we have split the correlation function into oscillatory and non-oscillatory contributions. The GUE result is proportional to $\frac{1}{\epsilon^2}$ whereas the GOE result leads to an infinite asymptotic power series in $\frac{1}{\epsilon}$.
The terms involving odd powers of $i$ are meaningful if $\epsilon$ is taken with a small positive imaginary part.
Keating and Robbins used the so-called diagonal approximation to recover the leading non-oscillatory term proportional to $\frac{1}{\epsilon^2}$ (corresponding to the linear term in the Fourier transform).
Depending on the time reversal properties of the system {\it as well as the type of the representation}
they obtained
$-\frac{1}{2\epsilon^2}$ as predicted by the GUE or $-\frac{1}{\epsilon^2}$ as predicted by the GOE.

In the present paper we show this agreement with RMT carries over to the full correlation function by building on recent results for systems without geometrical symmetries.
The starting point for all such work is the Gutzwiller trace formula \cite{Gut90}
\begin{equation}\label{Gutzwiller_Formula}
\rho(E) \sim \bar{\rho}(E) + \frac{1}{\pi\hbar}\mbox{Re}\sum_aT_aF_ae^{iS_a(E)/\hbar}
\end{equation}
which relates the quantum density of states to a sum over the actions $S_a(E)$ of the periodic orbits of the corresponding classical system, weighted by the primitive period $T_a$ and complex stability amplitude $F_a$ (which contains the Maslov index). Convergence can be ensured by taking the energy $E$ with a large enough imaginary part which limits the contributions from long periodic orbits. By directly inserting the trace formula into (\ref{Correlation_Function}) one sees that spectral correlations are determined by correlations between periodic orbits \cite{Arg93}.
The evaluation of these correlations allows one to obtain all non-oscillatory contributions to $R(\epsilon)$. The first term is given by the aforementioned diagonal approximation \cite{Han84, Ber85} taking into account pairs of identical or (in time-reversal invariant systems) mutually time-reversed orbits. The remaining terms arise from pairs of periodic orbits that differ by their connections inside close `encounters' \cite{Sie01,Sie02,Mul04,Mul05,Mul06}. For example, one orbit could contain a self crossing with a small intersection angle while the other orbit narrowly avoids the crossing. By analysing the nature of such encounters one can also show why for broken time-reversal invariant systems there are no further non-oscillatory terms beyond the diagonal approximation.

Currently our knowledge of orbit correlations is not sufficient to obtain the oscillatory components directly, instead one must take an indirect approach by relating long and short orbits through a `resummation' or `bootstrapping' procedure. This technique can be applied directly to the periodic orbit sum (\ref{Gutzwiller_Formula}) \cite{Bog96a}, however it turned out more convenient to consider the resummation of pseudo-orbits (collections of periodic orbits) instead. The resulting Riemann-Siegel lookalike formula \cite{Ber90,Kea92,Ber92} manifestly preserves the unitarity of the quantum time evolution and was combined with a so called generating function \cite{Heu07} in order to access the oscillatory contributions to the two-point correlation function \cite{Kea07}. Inspired by field-theoretic approaches the authors of \cite{Mul09} were then able to relate contributions from sets of pseudo orbits with small action differences to terms arising in the nonlinear sigma model. This resulted in the reproduction of  the full GOE and GUE expansions for systems with and without time reversal invariance.

We will combine this approach with results from representation theory to investigate systems
with discrete geometrical symmetries.
Our techniques have interesting analogies to the work of Boris Gutkin which recovers the non-oscillatory contributions to $R(\epsilon)$ for  ``cellular'' billiards \cite{Gut11}. Correlations between different subspectra are also considered in \cite{Bra11}.

The paper is organized as follows. In Section \ref{Symmetry_in_QM} we will outline some of the basic principles behind the application of representation theory in quantum mechanics.
In Section  \ref{Sym_Proj_GF} we will then introduce a generating function determining the correlations
inside each subspectrum.
This generating function and the corresponding correlation function will be evaluated
in Sections \ref{Diagonal_Approx} and \ref{Off-dia_contr}
using correlations between orbits as well as results from representation theory.
In Section \ref{Full_corr_function} we will generalize to cross-correlation functions between different subspectra as well as the full correlation function of a symmetric quantum system. Conclusions will be given in Section \ref{Conclusions}.

\section{Symmetry in quantum mechanics}\label{Symmetry_in_QM}

To get started, we give a brief outline of parts of representation theory that are helpful for understanding symmetries from a semiclassical point of view. For details see e.g. \cite{Ell79,Ham62,Wig59,Cor97}.

\begin{figure}[ht]
\centerline{\includegraphics[width=0.25\paperwidth]{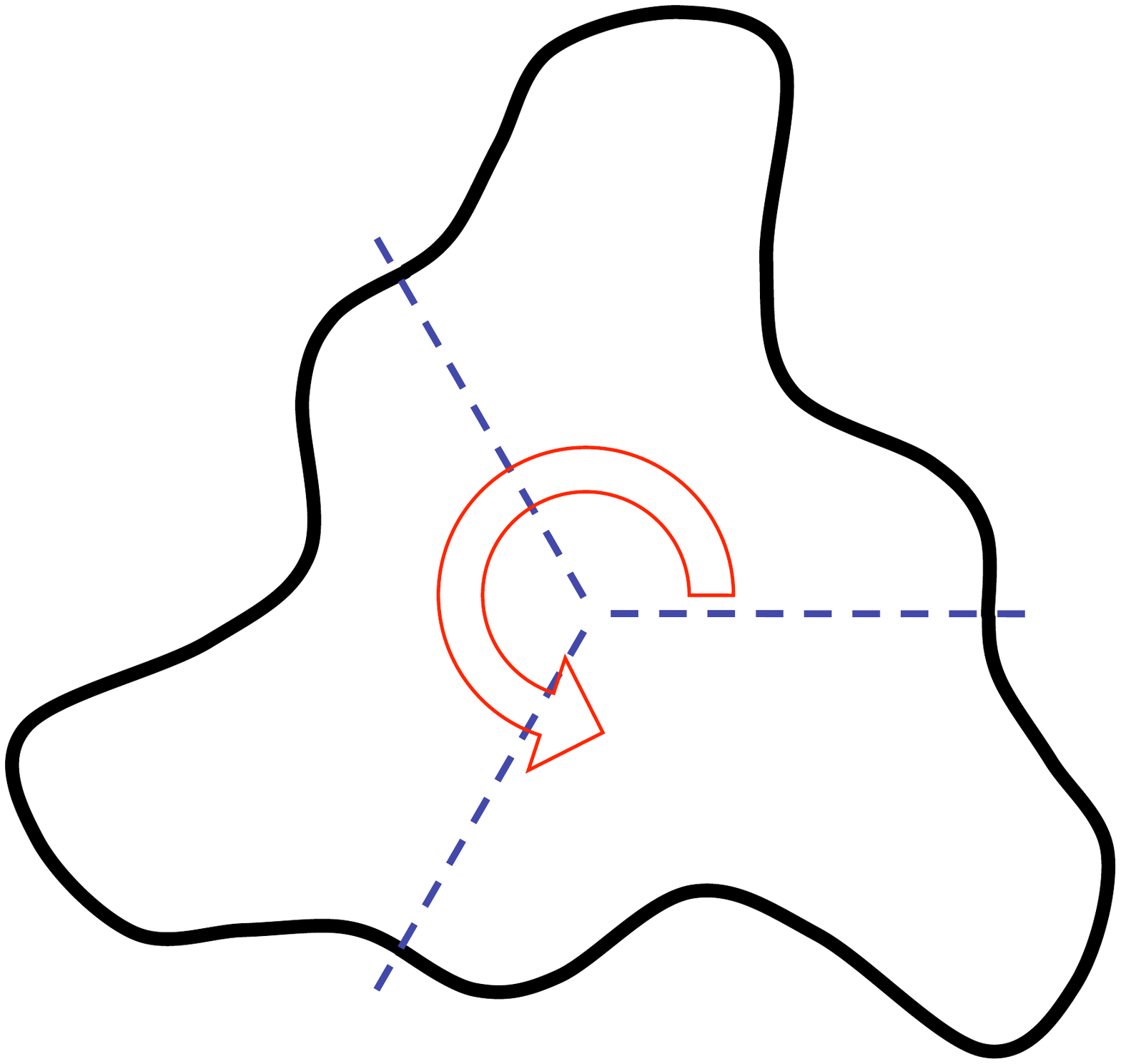}}
\caption{A billiard system that is invariant under rotations by $2 \pi/3$.}
\label{billiard}
\end{figure}

{\it Classical symmetries} are characterized by a group $G$ of symmetry operations that leave the system invariant.
For example if a classical Hamiltonian $H(q,p)$ is invariant under rotations by $2\pi/3$ the symmetry group is given
by $G=C_3 = \{e,g,g^2\}$, where $g$ denotes  rotation by $\frac{2\pi}{3}$ and $e$ is the identity. Figure \ref{billiard}
shows a billiard system with this symmetry. 

If the group $G$ consists of point transformations then in {\it quantum mechanics} it induces transformations of wave functions
through the definition $U(g)\psi(\boldsymbol r)=\psi(g^{-1}(\boldsymbol r))$, $g \in G$. The unitary operators $U(g)$ form a {\it representation}
of the group $G$ because they satisfy $U(g_2) \, U(g_1)=U(g_2g_1)$ for all $g_1,g_2\in G$. In a symmetric system the transformations
$U(g)$ commute with the Hamiltonian, $[U(g),H] = 0$ for all $g \in G$, and $G$ is said to be a symmetry group of the Hamiltonian.
One can consider also more general symmetry groups for which the operators $U(g)$ are not related to coordinate transformations
(for example, the Hecht Hamiltonian in \cite{Rob89} has a symmetry in angular momentum space).

In a quantum system with a discrete symmetry group one can split the spectrum into appropriate subspectra. For example,
if the symmetry group is $C_3$ then one can define three subspaces by requiring that the wave functions are periodic
with respect to rotations by $2 \pi/3$, up to a phase factor whose third power is one: 
$\psi(r,\phi-2\pi/3) = e^{2\pi i \alpha/3} \psi(r,\phi)$ where $\alpha = 0,1,2$. More formally, it follows from $[U(g), H] = 0$ that one can find a basis in which the action of $U(g)$ on a particular basis function is determined by the irreducible unitary matrix representation $M^{(\alpha)}(g)$ (see \cite{Ell79}-\cite{Cor97}) as follows
\begin{equation}\label{transf1}
U(g)\left| \alpha,i,n\right> = \sum_{j=1}^{s_{\alpha}}M^{(\alpha)}_{ji}(g) \left| \alpha,j,n\right> , \qquad
H \left| \alpha,i,n\right> = E^{(\alpha)}_n \left| \alpha,i,n\right> .
\end{equation}
Here $\alpha$ labels the different irreducible representations, $i$ the components within this representation, $n$ labels basis functions with the same $\alpha$ and $i$, and $s_\alpha$ denotes the dimension of the representation. Hence when $U(g)$ acts on the entire Hilbert space it is block-diagonal, with each block an $s_{\alpha} \times s_{\alpha}$ matrix given by $M^{(\alpha)}(g)$. Moreover the $s_\alpha$ eigenfunctions that correspond to this block all have the same energy and so the eigenvalue $E^{(\alpha)}_n$ is $s_\alpha$-fold degenerate. One can hence decompose the spectrum of a symmetric system into subspectra that correspond to the different irreducible representations of the symmetry group.

Note that the irreducible representations of a group satisfy $\sum_{\alpha} s_{\alpha}^2 = |G|$
where $|G|$ is the number of elements in $G$. In the example above the symmetry group $C_3$
has three one-dimensional representations, labelled by $\alpha=0,1,2$, and given by the one-dimensional
matrices $M^{(\alpha)}(g^N)=e^{2\pi i N \alpha/3}$ for $N=0,1,2$. Hence the eigenfunctions of the Hamiltonian satisfy
\begin{equation}
U(g^N) \psi^{(\alpha)}_n(r,\phi) = \psi^{(\alpha)}_n(r,\phi-2\pi N/3) = e^{2\pi i N \alpha/3}\psi^{(\alpha)}_n(r,\phi),
\end{equation}
in agreement with the more informal approach above. For completeness we note that the projection operator onto the
subspace $\alpha$ can be explicitly given in the form \cite{Ell79,Ham62}
\begin{equation}\label{projform}
{P}_{\alpha} = \frac{s_{\alpha}}{|G|} \sum_{g}  U^{\dagger}(g) \, {\rm tr} M^{(\alpha)}(g) \, ,
\end{equation}
satisfying $P_{\alpha} \left| \alpha',i,n\right> = \delta_{\alpha \alpha'} \left| \alpha',i,n\right>$ and $P_{\alpha}^2=P_{\alpha}$.

In the following it will be important that one can distinguish between three different types of representations.
An irreducible representation is {\it real} if the matrices $M^{(\alpha)}(g)$ are real, or can be made real by a
simultaneous similarity transformation\footnote{Irreducible representations that are related by a
simultaneous similarity transformation $\tilde{M}^{(\alpha)}(g) = S M^{(\alpha)}(g) S^{-1}$, $g \in G$, are equivalent.}.
The representation is {\it pseudo-real} if it is not real but all matrices $M^{(\alpha)}(g)$ have real traces.
In the remaining case where not all matrices have real traces the representation is {\it complex}.
Complex representations come in complex conjugate pairs, because for every complex representation $M^{(\alpha)}(g)$ there exists a complex conjugate representation
$M^{(\beta)}(g)=M^{(\alpha)}(g)^*$, which is inequivalent to $M^{(\alpha)}(g)$. If a quantum system is invariant
under complex conjugation, which corresponds to the time reversal operation for non-spin systems,
this leads to an additional degeneracy between the spectra of complex conjugate representations.
For example, for a time-reversal invariant system with symmetry group $C_3$ one can use complex
conjugation to obtain from every solution that satisfies $ \psi(r,\phi-2\pi/3) = e^{2\pi i/3} \psi(r,\phi)$ another
solution that satisfies $ \psi(r,\phi-2\pi/3)^* = e^{- 2\pi i/3} \psi(r,\phi)^* = e^{4\pi i/3} \psi(r,\phi)^*$ and has the same energy. For real and
pseudo-real representations the complex conjugate representation is equivalent to the original one\footnote{An
additional two-fold degeneracy is also present in pseudo-real representations if the quantum system is invariant
under complex conjugation due to a different mechanism \cite{Wig59,Cor97}.}.

An important property of the representation matrices $M^{(\alpha)}(g)$ that we will use in the following
is the group orthogonality relation that follows from Schur's lemmas. It states that
\begin{equation} \label{grouportho}
\frac{1}{|G|} \sum_g M_{ij}^\alpha(g) \, M_{kl}^\beta(g)^* = \frac{\delta_{\alpha \beta} \, \delta_{i k} \, \delta_{j l}}{s_\alpha} \, .
\end{equation}
In the semiclassical approach the representation matrices usually contribute through their traces
$\chi_\alpha(g)={\rm tr}\,M^{(\alpha)}(g)$, called {\it characters}. By taking traces in (\ref{grouportho})
one obtains the orthogonality relation for characters
\begin{equation}\label{orthcond2}
\frac{1}{|G|} \sum_{g\in G} \chi_{\alpha}(g)\chi^*_{\beta}(g)  = \delta_{\alpha\beta} \, .
\end{equation}
We will also need the following identities for group averages of characters
\begin{eqnarray}
\label{corr1}
\frac{1}{|G|}\sum_{g \in {G}}\chi_{\alpha}(agbg) &=& \frac{c_{\alpha}}{s_{\alpha}}\chi_{\alpha}(ab^{-1}) \, , \\
\label{corr2}
\frac{1}{|G|}\sum_{g \in {G}}\chi_{\alpha}(ag)\chi_{\alpha}(bg^{-1}) &=& \frac{1}{s_{\alpha}}\chi_{\alpha}(ab) \, , \\
\label{corr3}
\frac{1}{|G|}\sum_{g \in {G}}\chi_{\alpha}(agbg^{-1}) &=& \frac{1}{s_{\alpha}}\chi_{\alpha}(a)\chi_{\alpha}(b) \, .
\end{eqnarray}
Here $a$ and $b$ are arbitrary group elements. In the first identity we respectively have $c_\alpha=1,0,-1$
for real, complex, and pseudo-real representations. The second and third identity hold regardless of the type
of representation. These identities are proved in the appendix of \cite{Bol06}. They can be obtained by
manipulating a simpler identity from \cite{Ham62} and the group orthogonality relation (\ref{grouportho}).

In this article we will show that for spinless time reversal invariant systems the subspaces associated to real representations
obey GOE statistics, and subspaces associated to complex representations obey GUE statistics in spite of time reversal invariance.
Pseudo-real representations lead to GSE statistics \cite{Joy11} but will be excluded in the present paper.
For systems without time reversal invariance all subspaces obey GUE statistics.
Different subspectra are uncorrelated apart from the degeneracies mentioned above that are related to
time-reversal invariance.

\section{Generating Function}

\label{Sym_Proj_GF}

We are interested in the statistics of the energy levels $E_n^{(\alpha)}$ in a subspace $\alpha$, where the $s_{\alpha}$-fold degeneracy has been removed. The subsequent level density $\rho_\alpha(E)=\sum_n\delta(E-E_n^{(\alpha)})$ can be accessed from the following trace
\begin{equation}
\rho_\alpha(E)=-\frac{1}{\pi}\lim_{E^+ \to E}{\rm Im}\sum_n\frac{1}{E^+ - E_n^{(\alpha)}} = -\frac{1}{\pi}\lim_{E^+\to E}{\rm Im}\left(\frac{1}{s_\alpha}{\rm tr}\,P_\alpha\frac{1}{E^+ -H}\right),
\end{equation}
where $P_\alpha$ (see Eq. (\ref{projform})) is the projector onto the subspace $\alpha$ and $E^+ = E +i\eta$ has a small positive imaginary part that is taken to zero. Similarly the correlation function
\begin{equation}
R_\alpha(\epsilon)=\frac{1}{{\bar\rho_\alpha}^2} \left< \rho_\alpha\left(E + \frac{\epsilon}{2\pi \bar{\rho}_\alpha} \right) \rho_\alpha\left(E - \frac{\epsilon}{2\pi \bar{\rho}_\alpha} \right) \right>_E -1
 \end{equation} can be obtained as the limit for $\epsilon^+ = \epsilon + i\eta \to \epsilon$ of the real part of the complex correlator
\begin{equation}
\label{C}
C_\alpha(\epsilon^+)=\frac{1}{2\pi^2\bar\rho_\alpha^2}\left\langle\;
\frac{1}{s_\alpha}{\rm tr}\,P_\alpha\frac{1}{E+\frac{\epsilon^+}{2\pi\bar{\rho}_\alpha}-H}\;\;
\frac{1}{s_\alpha}{\rm tr}\,P_\alpha\frac{1}{E-\frac{\epsilon^+}{2\pi\bar{\rho}_\alpha}-H}\;
\right\rangle_E - \frac{1}{2}\;.
\end{equation}
Here $\bar{\rho}_\alpha$ is the mean level density of the subspectrum. It is related to that of the full spectrum by
$\bar{\rho}_\alpha \sim s_\alpha \bar{\rho}/|G|$ \cite{Lau95}\footnote{Here we have translated the result 
of \cite{Lau95} to our notation. In contrast to \cite{Lau95} our $\bar\rho_\alpha$ contains each degenerate level only once.}.
The random matrix predictions for the complex correlation function are given by Eqs. (\ref{GUE_Expansion}) and
(\ref{GOE_Expansion}), in each case dropping the restriction to the real part.

To access the level density using Riemann-Siegel resummation we must first relate the trace of the symmetry projected
resolvent to the corresponding spectral determinant $\Delta_\alpha(E)$  that vanishes at the energies $E_n^{(\alpha)}$.
The connection is given by the usual relation that the trace of the resolvent is the logarithmic derivative of the spectral
determinant. Hence we have\footnote{In general, the trace of the resolvent and the
determinant need to be regularised,
see \cite{Vor87}. This leads to a regularisation dependent prefactor of the determinant that it is not relevant for the following
semiclassical calculations.}
\begin{equation}\label{Green_func_rel}
\Delta_{\alpha}(E^+) \propto \exp\left(\int^{E^+} \frac{1}{s_{\alpha}}\mbox{tr}\left[{P}_{\alpha}\frac{1}{E' - H}\right] dE' \right),
\end{equation}
and vice versa
\begin{equation}\label{density_of_states}
\frac{1}{s_\alpha}{\rm tr}\,P_\alpha\frac{1}{E-H}
=-\left.\frac{\partial}{\partial E'}\frac{\Delta_\alpha(E)}{\Delta_\alpha(E')}\right|_{E'=E}\;.
\end{equation}
If we use this idea for both traces in (\ref{C}) we can then access the complex correlator by
\begin{equation}\label{Full_Gen_Func_Relation}
C_{\alpha}(\epsilon^+) = -\left.2\frac{\partial^2 Z_{\alpha}}{\partial\epsilon_A\partial\epsilon_B}\right|_{(\|)} - \frac{1}{2}
\end{equation}
where the generating function $Z_\alpha$ is defined by
\begin{equation}\label{Generating_Function}
Z_{\alpha}(\epsilon_A,\epsilon_B,\epsilon_C,\epsilon_D) = \left<
\frac{\Delta_{\alpha}(E + \epsilon_C/2\pi\bar{\rho}_{\alpha})\Delta_{\alpha}(E - \epsilon_D/2\pi\bar{\rho}_{\alpha})}
{\Delta_{\alpha}(E +\epsilon_A/2\pi\bar{\rho}_{\alpha})\Delta_{\alpha}(E - \epsilon_B/2\pi\bar{\rho}_{\alpha})}\right>_E,
\end{equation}
and $(\|)$ denotes the matching conditions $\epsilon_A,\epsilon_B,\epsilon_C,\epsilon_D = \epsilon^+$.

We now want to derive a semiclassical approximation for the spectral determinant and thus for $Z_\alpha$. For this purpose we need to find a semiclassical expression for the trace of the symmetry projected Green's function. Fortunately this has already been obtained by Robbins \cite{Rob89} (see also Seligmann and Weidenm\"uller \cite{Sel93}) and is given by
\begin{equation}\label{Robbins_Forumla}
\frac{1}{s_\alpha}\mbox{tr} \left[{P}_{\alpha} \left(\frac{1}{E^+ - H}\right)\right] \sim \bar{g}_{\alpha}(E^+) - 
\frac{i}{\hbar}\sum_a T_a F_a \chi_{\alpha}(g_a) e^{iS_a(E^+)/\hbar}.
\end{equation}
Here $\bar{g}_\alpha(E^+)$ is the smooth part of the trace with the $1/s_\alpha$ prefactor. It is related to the mean density of states by
$\mbox{Im} \, \bar{g}_\alpha(E^+) = -\pi \bar{\rho}_{\alpha}(E^+)$.
Expression (\ref{Robbins_Forumla}) contains the same primitive orbit period $T_a$, action $S_a$ and stability amplitude $F_a$  as the usual Gutzwiller formula (\ref{Gutzwiller_Formula}), however now the sum is over all orbits $a$ which are periodic in the {\it fundamental domain}\footnote{We omit the contribution from orbits confined to the boundary of the fundamental domain. These orbits involve multiple repetitions of a primitive periodic orbit. Due to the exponential proliferation of orbits with increasing period their contribution is negligible.} but not necessarily periodic when unfolded to the full system. However their final point is always related to the initial point via one of the symmetry operations $g_a$. The weight then contains the character $\chi_\alpha(g_a)$ that corresponds to this symmetry operation in the representation $\alpha$.

\begin{figure}[ht]
\centerline{
\includegraphics[width=0.35\paperwidth]{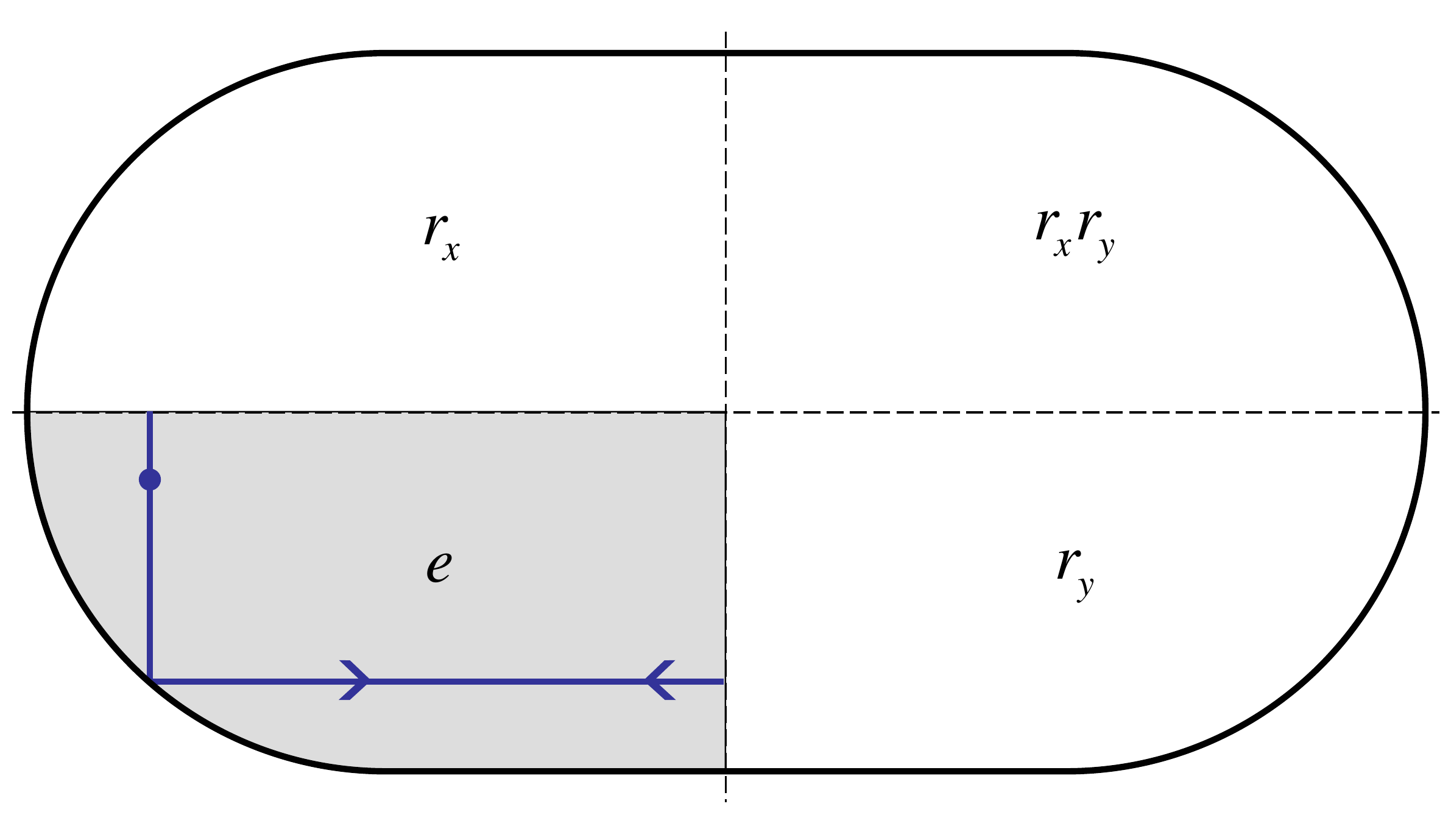}
\hspace{10pt}
\includegraphics[width=0.35\paperwidth]{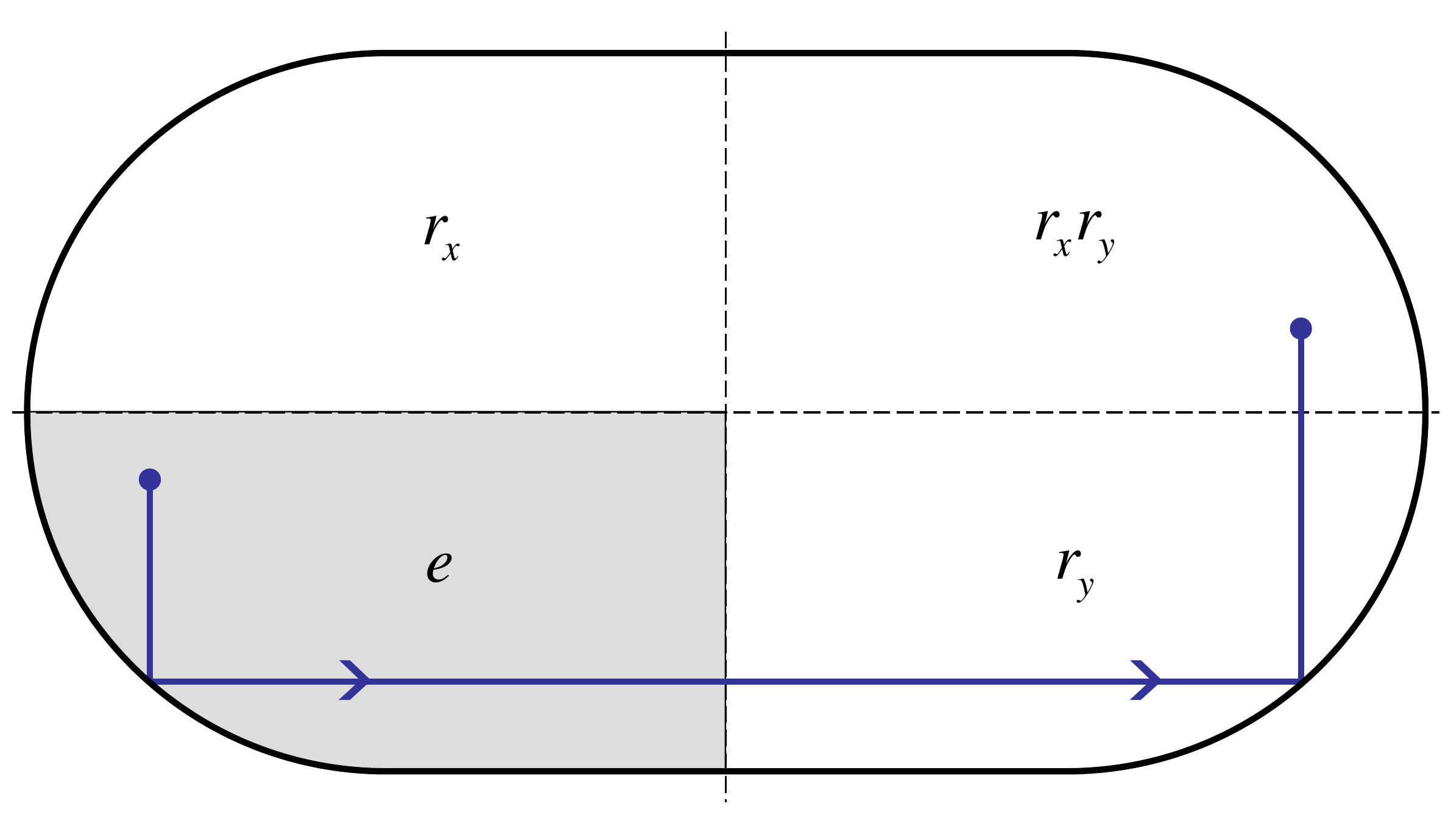}}
\caption{(a) A self-retracing periodic orbit in the fundamental domain is unfolded to (b) a partial orbit in the full stadium.}
\label{stadium}
\end{figure}

This may be illustrated using the stadium billiard which has a four-fold symmetry given by the group $S_2 \otimes S_2 = \{e,r_x,r_y,r_xr_y\}$, generated by the two reflection operators $r_x$ and $r_y$. The fundamental domain consists of a quarter of the stadium with specular reflection conditions at its boundary (shown in grey in Figure \ref{stadium}). Periodic orbits in the fundamental domain may be unfolded to the full system as shown in Figure~\ref{stadium}. In this example, a periodic orbit that retraces itself in the fundamental domain where it strikes both the $x$ and the $y$ axis (Figure \ref{stadium}(a)) will traverse both these axes in the full system and finish in the domain associated with the $r_xr_y$ group element (Figure \ref{stadium}(b)). Hence the corresponding weight is $\chi_\alpha(r_xr_y)$. We may view the group generators as symbols, allowing each section of an orbit to be assigned a particular symbol sequence and therefore a particular group element comprised of these generators.

If we insert (\ref{Robbins_Forumla}) into (\ref{Green_func_rel}) we obtain the following semiclassical expression for our spectral determinant\footnote{We neglect in the following the contributions of non-primitive periodic orbits to the trace formula since they are exponentially suppressed in the limit of long periods. Furthermore, the real part of $\bar{g}_\alpha$
is regularisation dependent and included in the proportionality factor.}
\begin{eqnarray}\label{scspecdet}
\Delta_{\alpha}(E^+) & \propto & \exp \left(-i\pi\bar{N}_{\alpha}(E^+) - \sum_a F_a \chi_{\alpha}(g_a)e^{iS_a(E^+)/\hbar}\right) \nonumber \\
& = & e^{-i\pi\bar{N}_{\alpha}(E^+)}\sum_A (-1)^{n_A}G^{(\alpha)}_A e^{iS_A(E^+)/\hbar}
\end{eqnarray}
where $\bar N_\alpha(E)$ is the mean counting function obtained by integrating $\bar\rho_\alpha(E)$. The sum is now over pseudo-orbits (finite collections of periodic orbits) $A$ with cumulative actions $S_A$ and amplitude factors $G^{(\alpha)}_A$, given as follows
\begin{equation}
G^{(\alpha)}_A = \prod_a \frac{(F_a \chi_{\alpha}(g_a))^{n_a}}{n_a!} \, , \hspace{20pt} S_A = \sum_a n_aS_a \, , 
\hspace{20pt} n_A = \sum_a n_a.
\end{equation}
Each $n_a$ runs from $0$ to $\infty$, but $n_A$ is always finite (and can be zero). For energies with negative imaginary parts one uses the relation
$\Delta_\alpha(E^-) = [\Delta_\alpha(E^+)]^*$. All information about our irreducible subspace is contained within the factor $G^{(\alpha)}_A$ and more specifically the characters within this factor. We also note the similarity between the expression (\ref{scspecdet}) and the symmetry reduced dynamical zeta functions defined in \cite{Lau91,Cvi93}. 

At this point our motivation for employing the spectral determinant becomes clear. As it stands, the expression (\ref{scspecdet}) suffers from exactly the same convergence issues as the Gutzwiller formula. However $\Delta_\alpha(E)$ obeys a functional equation which, via analytic continuation, can be used to show that truncating the sum in (\ref{scspecdet}) at half the (rescaled) Heisenberg time $T^{(\alpha)}_H = 2\pi\hbar\bar{\rho}_{\alpha}$ gives a contribution to $\Delta_\alpha(E)$ approximately equal to that of the complex conjugate of the remaining orbits \cite{Ber92}. In our context this resummation procedure, known as the `Riemann-Siegel lookalike' formula \cite{Ber90,Kea92} leads to an improved semiclassical approximation of the form
\begin{equation}\label{RS-lookalike}
\Delta_{\alpha}(E) = \sum_{T_A < T^{(\alpha)}_H/2}(-1)^{n_A}G^{(\alpha)}_A \exp\left(\frac{i}{\hbar} S_A(E) - i \pi\bar{N}_{\alpha}(E) \right) \hspace{10pt} + \hspace{10pt} \mbox{c.c.}
\end{equation}
where $T_A=dS_A/dE$ is the cumulative period of pseudo-orbits in $A$. In the semiclassical limit $T^{(\alpha)}_H \rightarrow \infty$ and one recovers all contributions. We now follow  \cite{Mul09} by inserting (\ref{RS-lookalike}) into the spectral determinants in the numerator of our generating function (\ref{Generating_Function}). This gives rise to four terms with oscillatory components in the numerators of the form $\exp(\pm i\pi[\bar{N}_{\alpha}(E + \epsilon_C/2\pi\bar{\rho}_{\alpha}) - \bar{N}_{\alpha}(E - \epsilon_D/2\pi\bar{\rho}_{\alpha})])$ and $\exp(\pm i\pi[\bar{N}_{\alpha}(E + \epsilon_C/2\pi\bar{\rho}_{\alpha}) + \bar{N}_{\alpha}(E - \epsilon_D/2\pi\bar{\rho}_{\alpha})])$. However those with additive phases oscillate rapidly and are assumed to give zero contribution after averaging over the energy E. Thus our semiclassical generating function $Z_{\alpha}$ may be written as the sum of two terms
\begin{equation}\label{SC_GF_Relation}
Z_{\alpha}(\epsilon_A,\epsilon_B,\epsilon_C,\epsilon_D) = Z^{(1)}_{\alpha}(\epsilon_A,\epsilon_B,\epsilon_C,\epsilon_D) + Z^{(2)}_{\alpha}(\epsilon_A,\epsilon_B,\epsilon_C^*,\epsilon_D^*)
\end{equation}
with the first given by
\begin{eqnarray}\label{genfunc1}
Z^{(1)}_{\alpha} & = & \left<
e^{i\pi\bar{N}_{\alpha}(E + \epsilon_A/2\pi\bar{\rho}_{\alpha})}\sum_A G^{(\alpha)}_A
e^{iS_A(E + \epsilon_A/2\pi\bar{\rho}_{\alpha})/\hbar}\right. \nonumber \\
& \times &
e^{-i\pi\bar{N}_{\alpha}(E - \epsilon_B/2\pi\bar{\rho}_{\alpha})}\sum_B G^{(\alpha)*}_B
e^{-iS_B(E - \epsilon_B/2\pi\bar{\rho}_{\alpha})/\hbar} \nonumber \\
& \times &
e^{-i\pi\bar{N}_{\alpha}(E + \epsilon_C/2\pi\bar{\rho}_{\alpha})}\sum_C (-1)^{n_C}G^{(\alpha)}_C
e^{iS_C(E + \epsilon_C/2\pi\bar{\rho}_{\alpha})/\hbar} \nonumber \\
& \times &
\left.e^{i\pi\bar{N}_{\alpha}(E - \epsilon_D/2\pi\bar{\rho}_{\alpha})}\sum_D (-1)^{n_D}G^{(\alpha)*}_D
e^{-iS_D(E - \epsilon_D/2\pi\bar{\rho}_{\alpha})/\hbar} \right>
\end{eqnarray}
and the second obtained by taking the complex conjugate of the third and fourth line, or equivalently
\begin{equation}\label{Gen_func_relation}
Z^{(2)}_{\alpha}(\epsilon_A,\epsilon_B,\epsilon_C^*,\epsilon_D^*) = Z^{(1)}_{\alpha}(\epsilon_A,\epsilon_B,-\epsilon_D^*,-\epsilon_C^*)\;.
\end{equation}
To simplify our expression we expand the mean counting function and action to obtain
\begin{equation}
\bar{N}_{\alpha}(E \pm \epsilon/2\pi\bar{\rho}_{\alpha}) \sim \bar{N}_{\alpha}(E) \pm\epsilon/2\pi
\end{equation}
and
\begin{equation}
S(E \pm \epsilon/2\pi\bar{\rho}_{\alpha})/\hbar \sim S(E)/\hbar \pm T(E) \epsilon/T^{(\alpha)}_H
\end{equation}
which are then inserted into the generating function to give the following expression
\begin{eqnarray}\label{SC_Generating_Function}
Z^{(1)}_{\alpha} & = & e^{i(\epsilon_A + \epsilon_B - \epsilon_C - \epsilon_D)/2}
\left<\sum_{A,B,C,D}G^{(\alpha)}_AG^{(\alpha)*}_BG^{(\alpha)}_CG^{(\alpha)*}_D(-1)^{n_C+n_D}\right.\nonumber \\
& \times & \left.e^{i(S_A - S_B + S_C - S_D)/\hbar}e^{i(T_A\epsilon_A + T_B\epsilon_B + T_C\epsilon_C + T_D\epsilon_D)/T^{(\alpha)}_H}\right>.
\end{eqnarray}
Contributions to the generating function will occur when the action difference  $\Delta S = S_A + S_C - S_B - S_D$ is
at most of the order of $\hbar$; summands for which the action difference is large in comparison to $\hbar$ are expected to wash out after averaging over the energy. Systematic contributions will obviously arise when the same orbits (modulo time-reversal) occur in both $A\cup C$ and $B\cup D$, extending Berry's original `diagonal approximation'. The remaining `off-diagonal' contributions will arise if the orbits in $B\cup D$ differ from those in $A\cup C$ only due to their connections inside encounters. Using these ideas the generating function of systems without symmetries was evaluated in \cite{Heu07,Kea07,Mul09}. The following sections extend this approach to systems with discrete symmetries by evaluating what effects the characters $\chi_{\alpha}(g_a)$, associated with each periodic orbit, have on these contributions.

\section{Diagonal Approximation}\label{Diagonal_Approx}

We start by evaluating the contribution from those orbits which have the same action in time-reversal invariant systems.
For this it helps to rewrite (\ref{SC_Generating_Function}) by converting each of the four pseudo-orbit sums back into an exponentiated sum over periodic orbits, i.e.
\begin{equation}
Z^{(1)}_{\alpha} = e^{i(\epsilon_A + \epsilon_B - \epsilon_C - \epsilon_D)/2}
\left< \exp \left(\sum_a F_a\chi_{\alpha}(g_a)e^{iS_a/\hbar}f^a_{AC}\right)
\exp \left(\sum_b F^*_b\chi^*_{\alpha}(g_b)e^{-iS_b/\hbar}f^b_{BD}\right) \right>
\end{equation}
with
\begin{equation}
f^a_{ij} = \left(e^{iT_a\epsilon_i/T^{(\alpha)}_H} - e^{iT_a\epsilon_j/T^{(\alpha)}_H}\right).
\end{equation}
The exponentials can then be re-expanded in terms of Taylor series to give
\begin{equation}
Z^{(1)}_{\alpha} = e^{i(\epsilon_A + \epsilon_B - \epsilon_C - \epsilon_D)/2}\left<
\sum_{n,m=0}^{\infty}\frac{1}{n!m!}\sum_{\substack{a_1,\ldots,a_n, \\ b_1,\ldots,b_m}} \prod_{i=1}^n
\prod_{j=1}^m F_{a_i}F^*_{b_j}\chi_{\alpha}(g_{a_i})\chi^*_{\alpha}(g_{b_j})e^{i(S_{a_i}-S_{b_j})/\hbar}f^{a_i}_{AC}f^{b_j}_{BD}\right>.
\end{equation}
To implement the diagonal approximation we assume that all contributions may be neglected unless the actions
satisfy $\sum_{i} S_{a_i} = \sum_{j} S_{b_j}$. For this to occur a periodic orbit $a_i$ must be matched to a periodic orbit $b_j$ that is either identical to $a_i$ or mutually time reversed\footnote{Again we neglect orbits that are just repetitions of shorter orbits. Furthermore, it is sufficient to consider only the case where all $a_i$ are different, because the other cases lead to negligible contributions.}, hence we can set $n = m$. Given that there are $n!$ ways of matching the $a_i$ to $b_j$ we can then cancel one of the factorials in the denominator. Furthermore, if the orbits are identical we can replace the character $\chi_\alpha^*(g_{b_j})$ by $\chi_\alpha^*(g_{a_i})$, whereas if they are mutually time reversed the corresponding group elements are mutually inverse and the characters are mutually adjoint (due to the unitarity of the representation). This allows us to replace  $\chi_\alpha^*(g_{b_j})$ by $\chi_\alpha^*(g_{a_i}^{-1})=\chi_\alpha(g_{a_i})$. The resulting expression can then be written in the form
\begin{equation}\label{Diag_approx01}
Z^{(1)}_{\alpha,\mbox{\scriptsize diag}} = e^{i(\epsilon_A + \epsilon_B - \epsilon_C - \epsilon_D)/2}\left[1 + \sum_{n=1}^{\infty}\frac{1}{n!}\sum_{a_1\ldots a_n} \prod_{i=1}^{n}|F_{a_i}|^2\kappa_{\alpha}(a_i)f^{a_i}_{AC}f^{a_i}_{BD}\right]
\end{equation}
with
\begin{equation}\label{diag_coeff}
\kappa_{\alpha}(a) = \chi_{\alpha}(g_a)[\chi^*_{\alpha}(g_a) + \chi_{\alpha}(g_a)];
\end{equation}
Note that $\kappa_{\alpha}(a)$ replaces the factor 2 for time-reversal invariant systems without symmetries.
Rearranging the sum and product in (\ref{Diag_approx01}) we recover the standard Taylor expansion in $n$ of an exponential yielding
\begin{equation}
Z^{(1)}_{\alpha,\mbox{\scriptsize diag}} =  e^{i(\epsilon_A + \epsilon_B - \epsilon_C - \epsilon_D)/2}
\exp \left(\sum_a \kappa_{\alpha}(a)|F_a|^2f^a_{AC}f^a_{BD}\right).
\end{equation}
Now following the arguments in \cite{Kea97}  we assume that the stability amplitudes $F_a$ and the group elements $g_a$
of the orbits are uncorrelated.
This allows $\kappa_{\alpha}(a)$ to be separated from the orbit sum and averaged independently over the group. 
Due to ergodicity this group average is \emph{uniform} as the probability of finding the end points of an unfolded orbit in any particular copy of the fundamental domain becomes uniform as $T \rightarrow \infty$. Therefore we may write
\begin{equation}\label{Diag_approx02}
Z^{(1)}_{\alpha,\mbox{\scriptsize diag}} = e^{i(\epsilon_A + \epsilon_B - \epsilon_C - \epsilon_D)/2}\exp\left(\sum_a|F_a|^2 f^a_{AC}f^a_{BD}\right)^{\kappa^{\mbox{\tiny diag}}_{\alpha}}
\end{equation}
where
\begin{equation}\label{diagaverage}
\kappa^{\mbox{\tiny diag}}_{\alpha} = \left<\chi_{\alpha}(g)[\chi^*_{\alpha}(g) + \chi_{\alpha}(g)]\right>_g =
\frac{1}{|G|}\sum_g |\chi_{\alpha}(g)|^2 + \frac{1}{|G|}\sum_g\chi_{\alpha}(g)^2,
\end{equation}
and the nature of the exponential allows us to change the factor of $\kappa^{\mbox{\tiny diag}}_{\alpha}$ to a power. In (\ref{diagaverage})
the first summand is simply 1 due to the orthogonality relation (\ref{orthcond2}) whereas the second summand depends on the type of representation. If $\alpha$ is real then the characters are real and we obtain the same result as for the first summand. However if $\alpha$ is complex then the complex conjugate of a character is the character of the complex conjugate representation and the second summand vanishes due to (\ref{orthcond2}). Hence
\begin{equation}
\kappa^{\mbox{\tiny diag}}_{\alpha} = \left\{\begin{array}{ll}
2 \hspace{10pt} & \alpha \hspace{5pt} \mbox{real} \\
1 & \alpha \hspace{5pt} \mbox{complex.}
\end{array}\right.
\end{equation}
Now, by using the Hannay and Ozorio de Almeida \cite{Han84} sum rule we can approximate the sum over periodic orbits in (\ref{Diag_approx02}), weighted by stability amplitudes $|F_a|^2$, by an integral of the form $\int_{T_0}^{\infty}\frac{dT}{T}$. Here $T_0$ is the minimum period from which the orbits tend to behave ergodically and after scaling with the Heisenberg time takes the lower limit of the integral $T_0/T^{(\alpha)}_H \rightarrow 0$ in the semiclassical limit. Computing this integral (see \cite{Mul09} for details) then leads to a semiclassical generating function
\begin{equation}\label{diagcom}
Z^{(1)}_{\alpha,\mbox{\scriptsize diag}} = e^{i(\epsilon_A + \epsilon_B - \epsilon_C - \epsilon_D)/2}
\left(\frac{(\epsilon_C + \epsilon_B)(\epsilon_A + \epsilon_D)}{(\epsilon_A + \epsilon_B)(\epsilon_C + \epsilon_D)}\right)^{\kappa^{\mbox{\tiny diag}}_{\alpha}},
\end{equation}
where $Z^{(2)}_{\alpha,\mbox{\scriptsize diag}}$ can be obtained from the relation (\ref{Gen_func_relation}). This allows us to obtain our complex correlator through differentiating our generating function as follows
\begin{equation}
C_{\alpha,\mbox{\scriptsize diag}}(\epsilon) =
-2\frac{\partial^2}{\partial\epsilon_A\partial\epsilon_B}
\left[Z^{(1)}_{\alpha,\mbox{\scriptsize diag}} + Z^{(2)}_{\alpha,\mbox{\scriptsize diag}}\right]_{(\|)}  - \frac{1}{2}.
\end{equation}
Finally, inserting (\ref{diagcom}) into the above relation when $\alpha$ is complex we achieve
\begin{equation}
C_{\alpha,\mbox{\scriptsize diag}}(\epsilon) = -\frac{1}{2\epsilon^2} + \frac{e^{2i\epsilon}}{2\epsilon^2}
\end{equation}
which is exactly the diagonal approximation predicted by RMT for the complex correlator in the GUE case (\ref{GUE_Expansion}). It is worth noting that this happens precisely because the second term in (\ref{diagaverage}), attributed to the correlation between a periodic orbit and its time-reversed partner, averages to zero. Hence, as remarked in \cite{Kea97} we observe in complex representations a mechanism akin to the introduction of an Aharonov-Bohm flux, in which time-reversed periodic orbits exist but their attributed phase factors average to zero \cite{Ber86a}. In contrast, for real subspaces the coefficient $\kappa^{\mbox{\tiny diag}}_{\alpha} =2$ means
\begin{equation}
C_{\alpha,\mbox{\scriptsize diag}}(\epsilon) = -\frac{1}{\epsilon^2}
\end{equation}
which corresponds to the diagonal term in the GOE complex correlation function.

Conversely, in systems where the classical time-reversal symmetry is broken those time-reversed periodic orbits are no longer available and so the coefficient (\ref{diagaverage}) becomes
\begin{equation}
\kappa^{\mbox{\tiny diag}}_{\alpha} = \left<\chi_{\alpha}(g)\chi^*_{\alpha}(g)\right>_g =  \left\{\begin{array}{ll}
1 \hspace{10pt} & \alpha \hspace{5pt} \mbox{real} \\
1 & \alpha \hspace{5pt} \mbox{complex,} \end{array}\right.
\end{equation}
and we obtain the GUE complex correlation function in both real and complex subspaces. In summary, with the diagonal approximation we showed agreement with the RMT predictions up to leading order $\epsilon^{-2}$, and it remains for us to validate all other orders.

\section{Off-Diagonal Contributions}\label{Off-dia_contr}

\begin{figure}[ht]
\centerline{
\includegraphics[width=0.25\paperwidth]{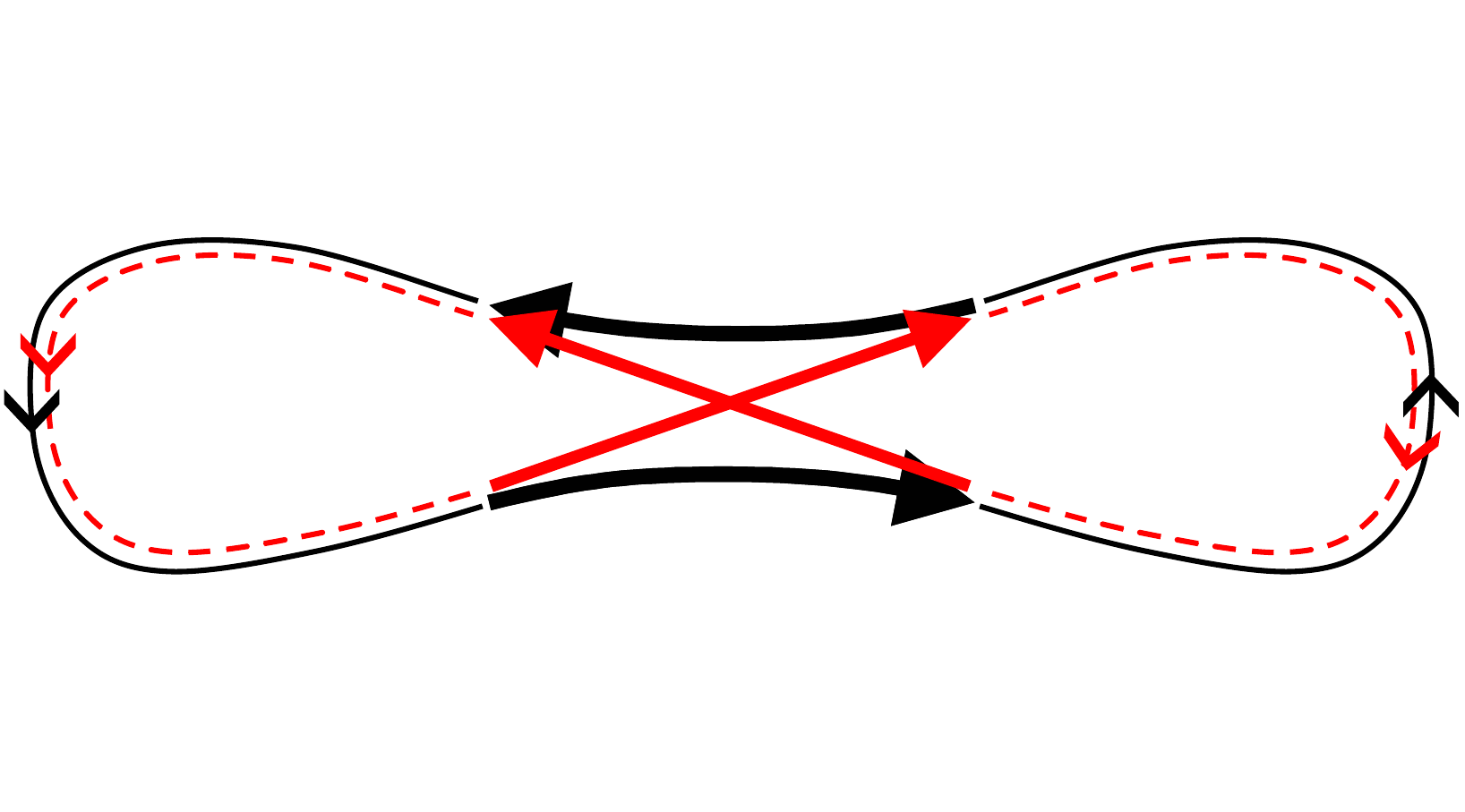}
\hspace{70pt}
\includegraphics[width=0.2\paperwidth]{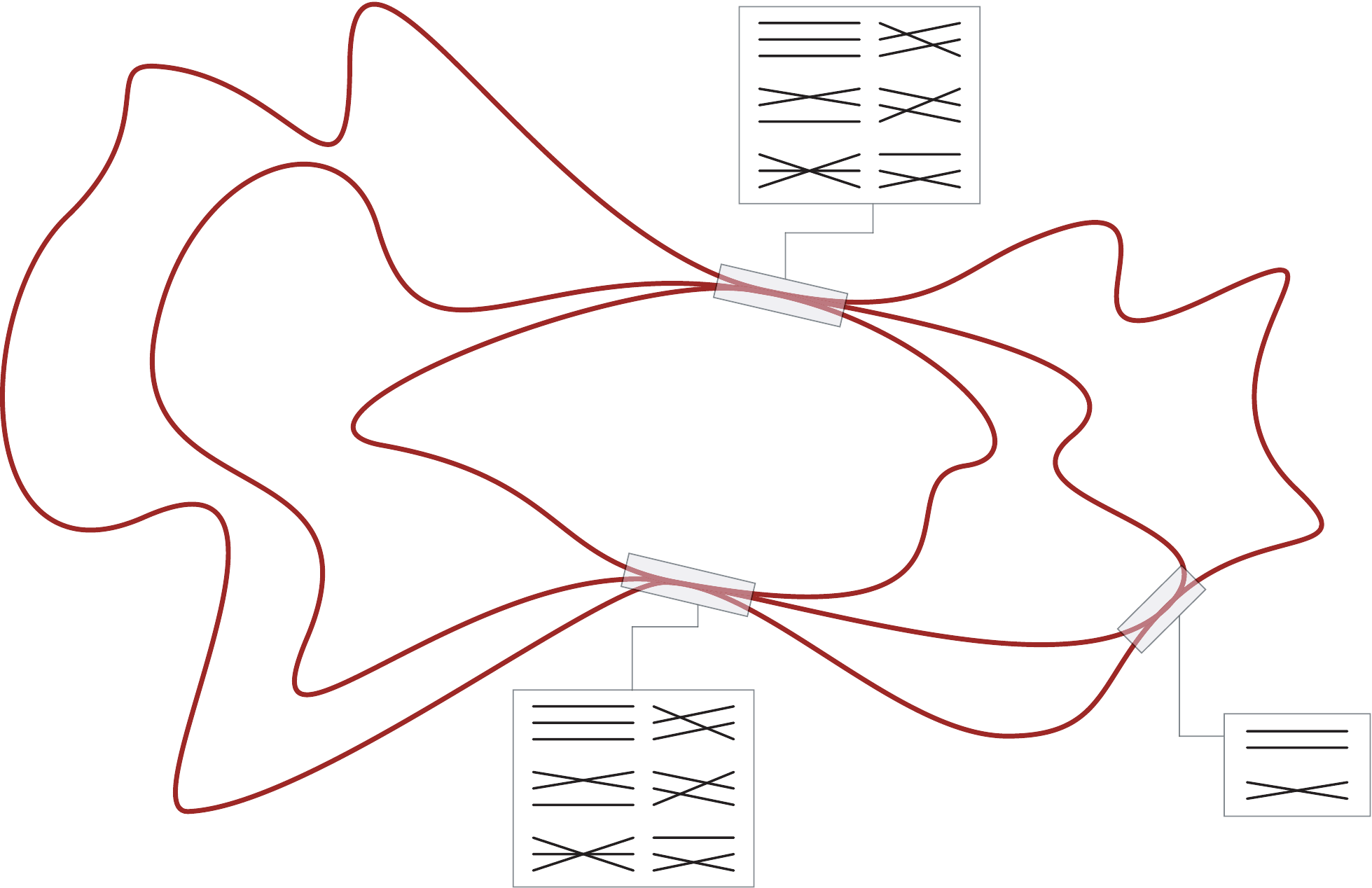}}
\caption{(a) A topological configuration space view of pair of periodic orbits with a single encounter. (b) Collections of bunched periodic orbits forming pseudo-orbits differing in multiple encounters (from \cite{Alt08}).}
\label{encounter}
\end{figure}

Further contributions to $Z^{(1)}_\alpha$ with small action differences arise if the orbits in $B\cup D$ closely follow those in $A\cup C$
  except in
so-called encounters where several orbit stretches  come close (see Figure \ref{encounter}). By changing the connections inside these encounters (which may lead to orbits splitting or merging), we can relate the orbits in $A \cup C$ to those in $B\cup D$. The possibility to switch connections in this way is a consequence of hyperbolicity.
In contrast the `links', i.e. the long parts of the orbits connecting the encounter stretches, are almost the same in $A\cup C$ and $B\cup D$ modulo time reversal.

We have to evaluate the contribution $Z^{(1)}_{\alpha,\mbox{\scriptsize off}}$
to the generating function (\ref{SC_Generating_Function}) arising
from $A\cup C$ and  $B\cup D$ differing by their connections in encounters, i.e.,
\begin{align}\label{Off_diag_GF}
Z^{(1)}_{\alpha,\mbox{\scriptsize off}} & = e^{i(\epsilon_A + \epsilon_B - \epsilon_C - \epsilon_D)/2}
\sum_{\substack{A,B,C,D \\ {\rm \;diff.\; in\; enc.}}}\left<G^{(\alpha)}_AG^{(\alpha)*}_BG^{(\alpha)}_CG^{(\alpha)*}_D(-1)^{n_C+n_D}
\right. \notag \\ & \qquad \qquad \qquad \left. \times
e^{i\Delta S/\hbar} \, e^{i(T_A\epsilon_A + T_B\epsilon_B + T_C\epsilon_C + T_D\epsilon_D)/T^{(\alpha)}_H}\right>,
\end{align}
where $\Delta S = S_A + S_C - S_B -S_D$.
To account for the possibility that some orbits in $B\cup D$ are obtained by changing connections in
$A\cup C$ whereas others are just identical  $Z^{(1)}_{\alpha,\mbox{\scriptsize off}}$  still has to be multiplied with $Z^{(1)}_{\alpha,\mbox{\scriptsize diag}}$.
Altogether we thus obtain the following expression for $Z^{(1)}_\alpha$
\begin{equation}
Z^{(1)}_{\alpha} = Z^{(1)}_{\alpha,\mbox{\scriptsize diag}}\left(1+Z^{(1)}_{\alpha,\mbox{\scriptsize off}}\right)\;.
\end{equation}
At present we only consider systems with time-reversal invariance. Systems without time-reversal invariance will be discussed in section \ref{nontime}.
The pseudo-orbit quadruplets that contribute to $Z^{(1)}_{\alpha,\mbox{\scriptsize off}}$ can have many different topologies. Following \cite{Mul09} (to which we refer for details) these topologies may be classified by using the notion of `structures' of quadruplets $(A,B,C,D)$. These structures are characterized by the number of encounters, the number of orbit stretches participating in each encounter, and the ordering of these stretches along the orbits. In systems with geometric symmetries, the contribution from each structure may be evaluated similarly as for the non-symmetric systems in \cite{Mul09} provided we account for the group characters associated to each periodic orbit and relevant factors of $|G|$ and $s_{\alpha}$ that arise from considering orbits in the fundamental domain. We discuss in the following the modifications of the calculations in \cite{Mul09} due to the presence of 
symmetries.

First we separate each pseudo-orbit of $A \cup C$ into the constituent periodic orbits $p_1,\ldots,p_n$ and each pseudo-orbit of $B \cup D$ into constituent periodic orbits $q_1,\ldots,q_m$. The combined amplitude factors from orbits in $A\cup C$ can be considered approximately equal to those in $B\cup D$. We then determine the action difference between both sets of periodic orbits which depends on the separation between the stretches involved in the encounters, with components $s$ and $u$ pointing
in the stable and unstable directions in phase space. We thus require a probability density $w_{T_1,T_2,\ldots T_n}(s,u)$ for finding in given periodic orbits ($p_1$ to $p_n$) $V$ encounters with a given structure and given $s$ and $u$. For systems without geometrical symmetries this density was evaluated using ergodicity and was shown to be proportional to $\Omega^{-(L-V)}$, where $\Omega$ is the volume of the energy shell and $L=\sum_{\sigma=1}^{V}l(\sigma)$ is the total number of encounter stretches (equivalently links), obtained from the number of stretches $l(\sigma)$ in each encounter $\sigma$. For systems with geometrical symmetries we must replace $\Omega$ with the volume of the fundamental domain $\Omega/|G|$ which leads to a factor of $|G|^{L-V}$ in comparison to the density in \cite{Mul09}. Summation over structures and integration over $s$ and $u$ therefore yields
\begin{eqnarray}\label{Z1alpha}
Z^{(1)}_{\alpha,\mbox{\scriptsize off}} & = & \sum_{\mbox{\tiny struct}} \frac{(-1)^{n_C+n_D}}{2^VV!\prod_{\sigma}l(\sigma)} \sum_{p_1,\ldots,p_n}
\kappa_{\alpha}(p_1,\ldots,q_m)\prod_{i=1}^n |F_{p_i}|^2 \nonumber \\
& \times &
\left<\int d^{L-V}s\hspace{2pt}d^{L-V}u \hspace{3pt}w_{T_1,\ldots,T_n}(s,u)e^{i\Delta S(s,u)/\hbar}e^{i(T_A\epsilon_A + T_B\epsilon_B + T_C\epsilon_C + T_D\epsilon_D)/T^{(\alpha)}_H}\right>.
\end{eqnarray}
Here the division by $2^VV!\prod_{\sigma}l(\sigma)$ avoids overcounting due to choices of $(A,B,C,D)$ that can be described in terms of several equivalent structures and the coefficient $\kappa_{\alpha}(p_1,\ldots,q_m)$ simply assembles the characters associated to each periodic orbit, i.e.
\begin{equation}
\kappa_{\alpha}(p_1,\ldots,q_m) = \chi_{\alpha}(g_{p_1})\ldots\chi_{\alpha}(g_{p_n})\chi^*_{\alpha}(g_{q_1})\ldots
\chi^*_{\alpha}(g_{q_m}).
\end{equation}
As in the previous section the crucial step then is to replace this coefficient by an average over all possible group elements for each periodic orbit
that are consistent with a considered structure. 
\begin{equation}
\kappa_{\alpha}^{\mbox{\tiny struct}} = \left<\chi_{\alpha}(g_{p_1})\ldots\chi_{\alpha}(g_{p_n})\chi^*_{\alpha}(g_{q_1})\ldots
\chi^*_{\alpha}(g_{q_m})\right>\,.
\end{equation}
We will discuss later in more detail how this average is performed for a given structure. Note that if one replaces a periodic orbit by its time-reversed equivalent then this corresponds to a different structure in the present formulation.

In summary, we have made the following changes compared to the non-symmetric case. In addition to the factor $\kappa_{\alpha}^{\mbox{\tiny struct}}$ and the factor of $|G|^{L-V}$ arising from the probability density $w_{T_1,T_2,\ldots T_n}(s,u)$ we have also replaced $T_H$ by $T_H^{(\alpha)}$ in the exponent of (\ref{Z1alpha}). This amounts to a multiplication of the exponent by $T_H/T_H^{(\alpha)}=\bar\rho/\bar\rho_\alpha=|G|/s_\alpha$. This can be taken into account by replacing all the $\epsilon_j$  by $\epsilon_j |G|/s_\alpha$ in the final result for the off-diagonal contribution to the generating function.

In the non-symmetric case the final result was \cite{Mul09}
\begin{equation}\label{Zoffdiag}
Z^{(1)}_{\mbox{\scriptsize off}}=\sum_{\rm struct} Z^{(1)}_{\mbox{\scriptsize off,\,struct}}\, , \quad \text{where} \quad
Z^{(1)}_{\mbox{\scriptsize off,\,struct}} =  \frac{(-1)^{n_C+n_D}}{2^VV!}
\frac{\prod_{\mbox{\tiny enc}} i(\epsilon_{A \hspace{1pt} \mbox{\scriptsize or} \hspace{1pt} C} +
\epsilon_{B \hspace{1pt} \mbox{\scriptsize or} \hspace{1pt} D})}{\prod_{\mbox{\tiny links}}
-i(\epsilon_{A \hspace{1pt} \mbox{\scriptsize or} \hspace{1pt} C} + \epsilon_{B \hspace{1pt} \mbox{\scriptsize or} \hspace{1pt} D})}\;.
\end{equation}
Here each link that is included in an original orbit $p_i$ and a partner orbit $q_j$ gives rise to a factor with subscripts depending on whether
$p_i$ belongs to $A$ or $C$ and whether  $q_j$ belongs to $B$ or $D$. The subscripts in the encounter factor depend on the orbits containing the beginning of a specified encounter stretch.

According to the discussion above we have the following modification in the symmetric case
\begin{equation}\label{sumkappa}
Z^{(1)}_{\alpha,\mbox{\scriptsize off}}=\sum_{\rm struct}\kappa_\alpha^{\rm struct} |G|^{L-V}\left(\frac{|G|}{s_\alpha}\right)^{V-L} \! \! Z^{(1)}_{\mbox{\scriptsize off,\,struct}}=
\sum_{\rm struct}\kappa_\alpha^{\rm struct} {s_\alpha}^{L-V}Z^{(1)}_{\mbox{\scriptsize off,\,struct}}
\end{equation}
where $Z^{(1)}_{\mbox{\scriptsize off,\,struct}}$ is given by (\ref{Zoffdiag}) and is symmetry independent.

As $Z^{(1)}_{\mbox{\scriptsize off,\,struct}}$ is known we only have to deal with
 $\kappa_\alpha^{\rm struct}$. We will see that for time-reversal invariant systems
$\kappa_\alpha^{\rm struct}$ makes sure that structures that actually require time reversal invariance are omitted if the representation is complex. In all other situations $\kappa_\alpha^{\rm struct}$ cancels
the factor ${s_\alpha}^{L-V}$.
\begin{equation}
\kappa_\alpha^{\rm struct}=\begin{cases}
s^{V-L}_{\alpha}& \text{if $\alpha$ real, or $\alpha$ complex and structure does not require time-reversal invariance,}\\
0 & \text{if $\alpha$ complex and structure does require time-reversal invariance.}
\end{cases}
\end{equation}

\subsection{Real Irreducible Representations}

\begin{figure}[ht]
\centerline{
\includegraphics[width=0.25\paperwidth]{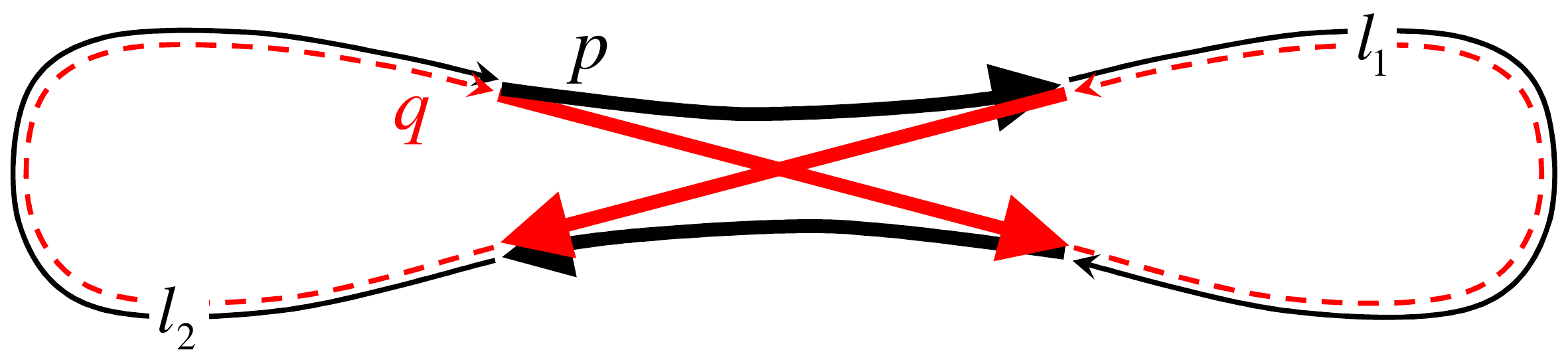}}
\caption{An orbit with two antiparallel encounter stretches, and its partner orbit obtained by changing connections inside the encounter.}
\label{pseudo_01}
\end{figure}

We start by evaluating $\kappa_{\alpha}^{\mbox{\tiny struct}}$ for real irreducible representations in time-reversal invariant systems.
To illustrate our approach we consider the orbit pairs introduced in \cite{Sie01} and depicted in Figure \ref{pseudo_01}.

Here $A\cup C$ only contains one periodic orbit $p$ where two encounter stretches are almost antiparallel. In the partner orbit $q$, included in $B\cup D$, the connections inside the encounter are changed such that one of the links, say the first, is reversed in time. Importantly, the group elements associated to $p$ and $q$ can be decomposed into the group elements associated to each link and encounter stretch. These elements indicate the symmetry operation that relates the copy of the fundamental domain in which the unfolded stretch or link starts to the copy in which it ends. If a stretch or link is changed slightly the group element stays the same and if it is reverted in time the group element is inverted. Thus in our example $p$ has a group element $g_p=l_2el_1e^{-1}$ (with $e$ and $l_i$ denoting encounter and link elements respectively)\footnote{e must not be confused with the identity element!} whereas in $q$ the first link is reverted in time such that $g_q=l_2 e l_1^{-1}e^{-1}$. The corresponding $\kappa_{\alpha}^{\mbox{\tiny struct}}$ is therefore
\begin{equation}
\kappa_{\alpha}^{\mbox{\tiny struct}}=\frac{1}{|G|^3}\sum_{e,l_1,l_2\in G}
\chi_\alpha(l_2 e l_1 e^{-1})\chi_\alpha^*(l_2 e l_1^{-1} e^{-1}).
\end{equation}
However the encounter elements can be removed as we can decompose $e$ as $e=\nu\eta$ and use the invariance of the group average under multiplication by an arbitrary group element to  substitute $\eta l_1 \eta^{-1} \rightarrow l_1$ and $\nu^{-1} l_2 \nu \rightarrow l_2$. This leads to
\begin{equation}
\label{srgroup}
\kappa_{\alpha}^{\mbox{\tiny struct}}
= \frac{1}{|G|^2}\sum_{l_1,l_2\in G}\chi_\alpha(l_2 l_1 )\chi_\alpha^*(l_2 l_1^{-1})
\end{equation}
which in the following sections we will show to yield $\frac{1}{s_{\alpha}}$.

In general for a particular structure $\kappa_{\alpha}^{\mbox{\tiny struct}}$ may be written as
\begin{equation}\label{grpcoeff}
\kappa_{\alpha}^{\mbox{\tiny struct}} = \left\langle\chi_{\alpha}(g_{p_1})\ldots\chi_{\alpha}(g_{p_n})\chi^*_{\alpha}(g_{q_1})\ldots\chi^*_{\alpha}(g_{q_m})\right\rangle_{e_1,\ldots,e_V,l_1,\ldots,l_L}.
\end{equation}
Here the group elements of the orbits $g_{p_1},\ldots,g_{p_n},g_{q_1},\ldots,g_{q_m}$
are alternating sequences of group elements $e_1,\ldots,e_V$ associated to the encounters (or their inverses)
and group elements $l_1,\ldots,l_L$ associated to the links (or their inverses). However, as in the example above we may split the encounters into the form $e_i = \nu_i\eta_i$ and redefine the link elements using the invariance of group sums under multiplication. Thus all encounter elements may be dropped from (\ref{grpcoeff}).

We now proceed to evaluate $\kappa_{\alpha}^{\mbox{\tiny struct}}$.
Bolte and Harrison have encountered a related situation \cite{Bol03,Bol06} on quantum graphs, where the group elements refer to the precession of spin around a periodic orbit, rather than unfolding periodic orbits in a system with a discrete symmetry. See also \cite{Heu01,Mul05} for the case of flows. In these instances group averages were performed for pairs of periodic orbits which determine the non-oscillatory contributions to the correlation function, without using the pseudo-orbit approach for the generating function. Here we extend these methods to discrete symmetries and structures involving arbitrarily many orbits.

\begin{figure}[ht]
\centerline{
\includegraphics[width=0.6\paperwidth]{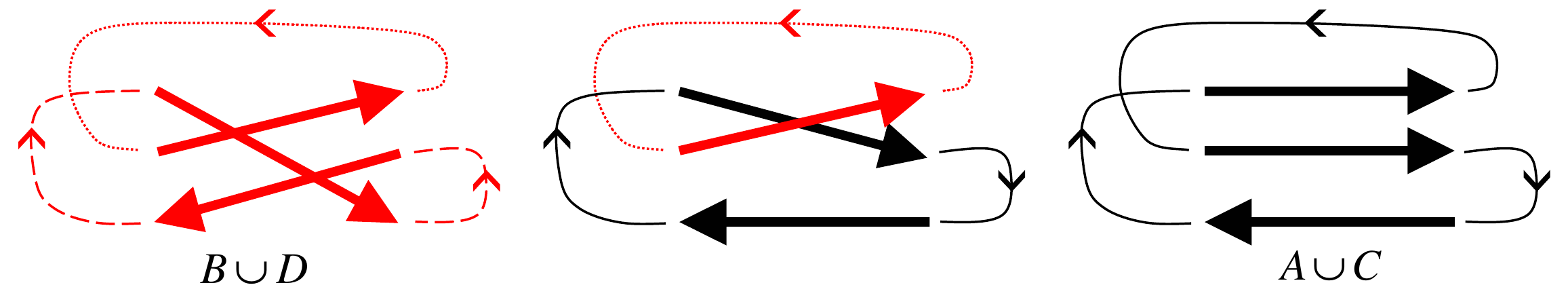}}
\caption{
For an encounter with three stretches
the connections can be changed in two steps each affecting only two stretches.}
\label{Permutation}
\end{figure}

We will use the fact that to fully change the connections in an encounter of $l$ stretches we need to perform $l-1$ steps that just interchange the connections between two encounter stretches. Therefore to fully change connections in all $V$ encounters and  change the orbits in $B \cup D$ to those contained in $A \cup C$ we need a total of  $\sum_{\sigma=1}^V(l(\sigma)-1)=L-V$ such steps. For example, Figure 4 depicts a structure involving $3-1=2$ steps. There are different types of steps, depending on whether the stretches involved are parallel or antiparallel and whether they belong to the same orbit or different orbits. We will show that regardless of these options
each step gives a factor of ${s_\alpha}^{-1}$, multiplication then leads to $s_\alpha^{-(L-V)}$.

\subsubsection{Anti-parallel encounter permutations} \label{AntiParallel_EP}

\begin{figure}[ht]
\centerline{
\includegraphics[width=0.7\paperwidth]{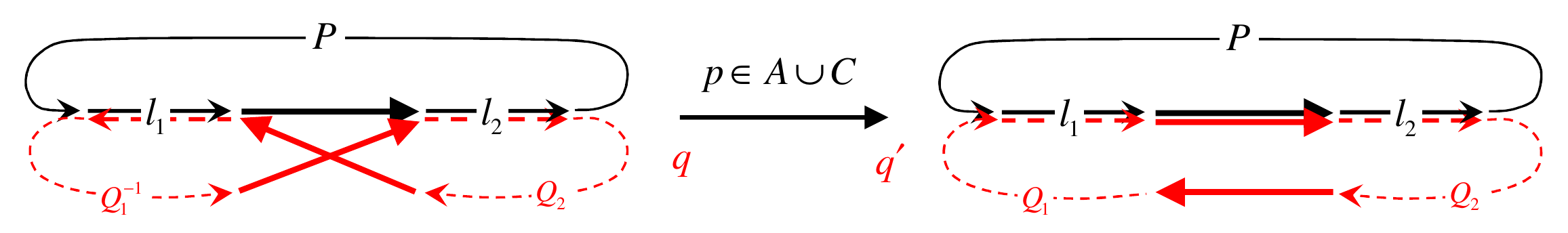}}
\caption{(a) Black/full line: an orbit $p \in A \cup C$ for which an encounter stretch (big arrow) that connects two links $l_1$ and $l_2$
has been highlighted. $P$ denotes the remaining part of the orbit. Red/dashed line: an orbit $q \in B \cup D$ that traverses the link $l_1$
in the opposite direction and $l_2$ in the same direction as $p$.
(b) Black/full line:  $p$  again. Red/dashed line: the orbit $q'$ where one of the reconnection steps leading from $q$ to $p$ has been performed.}
\label{Encs01}
\end{figure}

To derive the factor $s_\alpha^{-1}$ we compare the quantity $\kappa_{\alpha}^{\mbox{\tiny struct}}$ in Eq. (\ref{grpcoeff}) to the one for a different structure where one of the $L-V$ steps leading from $B\cup D$ to $A\cup C$ has been performed. Let us begin by selecting an individual encounter stretch of an orbit $p \in A \cup C$. This stretch connects two links whose group elements will be denoted by $l_1$ and $l_2$
(and its group element is absorbed in $l_1$ and $l_2$ as discussed above). Hence the group element associated to $p$ will be of the form
\begin{equation}\label{decomposition_11}
g_p=Pl_2l_1\;,
\end{equation}
where the product of group elements has to be read from right to left.
Here $P$ is associated to the remaining parts of $p$. Its precise form is unimportant for our calculation.
The links $l_1$ and $l_2$ also show up in $B\cup D$. We first consider the case where one of these links, say $l_1$, has been reverted in time and the two links form part of the same orbit $q\in B\cup D$. This situation is depicted in Fig.~\ref{Encs01}. The other cases are treated later. If we denote the group elements of the remaining parts of $q$ by $Q_1^{-1}$ and $Q_2$ then $q$ has a group element of the form
\begin{equation}\label{decomposition_12}
g_q=(Q_2l_2)(Q_1^{-1}l_1^{-1})\;.
\end{equation}
As illustrated by Fig.~\ref{Encs01}, the orbit $p$ contains an encounter stretch following $l_1$ and preceding $l_2$ (indicated by the big arrow).  The orbit $q$ has an almost parallel encounter stretch preceding the link $l_2$, and an almost antiparallel encounter stretch preceding the reverted link $l_1^{-1}$ (the two crossing big arrows). In the decomposition of $g_q$ in (\ref{decomposition_12}) the locations of these stretches are indicated by brackets.

Now we transform the structure to make $B\cup D$ more similar to $A\cup C$, by changing connections between the two antiparallel encounter stretches singled out above.
This replaces $q$ by an orbit $q'$ where one of the brackets in $g_q$, say $(Q_1^{-1}l_1^{-1})$, has been inverted, leading to
\begin{equation}
g_{q'}=(Q_2l_2)(l_1Q_1)\;.
\end{equation}
In $q'$ the links with group elements $l_1$ and $l_2$ follow each other directly, as in $p$. The structure in which $q'$ replaces $q$ is hence simpler than the original structure and now one less reconnection step is needed to go from $B\cup D$ to $A\cup C$.

We now want to relate the corresponding coefficients $\kappa_\alpha^{\mbox{\tiny struct}}$.
By inserting the orbit decompositions (\ref{decomposition_11}) and (\ref{decomposition_12})
for the original structure
 into (\ref{grpcoeff}) its coefficient can be represented as
\begin{equation}\label{coeff_sum}
\kappa_{\alpha}^{\mbox{\tiny struct}} = \left<\chi_{\alpha}(g_{p_1})\chi_{\alpha}(g_{p_2})\ldots\chi_{\alpha}(P l_2 l_1)
\chi^*_{\alpha}(g_{q_1})\chi^*_{\alpha}(g_{q_2})\ldots
\chi^*_{\alpha}((Q_2 l_2)(Q_1^{-1}l_1^{-1}))\right>
\end{equation}
with averages running over all  link elements.
In particular $\kappa_{\alpha}^{\mbox{\tiny struct}}$ involves the sum
\begin{equation}\label{coeff_sum2}
\frac{1}{|G|^2}\sum_{l_1,l_2\in G}\chi_{\alpha}(P l_2  l_1)\chi^*_{\alpha}((Q_2 l_2)(Q_1^{-1}l_1^{-1})).
\end{equation}
We now want to bring this sum to a form that contains $g_{q'}$ instead of $g_q$. For this purpose we use again that group averages are invariant under multiplication by an arbitrary member of $G$. This allows us to replace $l_1 \rightarrow x^{-1}l_1$ and $l_2 \rightarrow l_2 x$ and then average over $x$ without altering the value of the sum
\begin{equation}
\frac{1}{|G|^2}\sum_{l_1,l_2\in G}\frac{1}{|G|} \sum_{x\in G} \chi_{\alpha}(P l_2 l_1)\chi^*_{\alpha}((Q_2 l_2)x(Q_1^{-1}l_1^{-1})x).
\end{equation}
To evaluate the group average over $x$ we now use the identity
$\frac{1}{|G|}\sum_{g \in G}\chi_{\alpha}(agbg) = \frac{1}{s_{\alpha}}\chi_{\alpha}(ab^{-1})$
(see Eq. (\ref{corr1}) with $c_\alpha=1$ for real representations). This gives
\begin{equation}
\frac{1}{s_{\alpha}}\frac{1}{|G|^2}\sum_{l_1,l_2\in G} \chi_{\alpha}(P l_2 l_1)\chi^*_{\alpha}(Q_2l_2l_1 Q_1) = \frac{1}{s_{\alpha}}\frac{1}{|G|^2}\sum_{l_1,l_2\in G} \chi_{\alpha}(g_p)\chi^*_{\alpha}(g_{q'})\;.
\end{equation}
Hence by performing an antiparallel encounter exchange en route to changing  $B \cup D$ to  $A \cup C$ we
have expressed $\kappa_{\alpha}^{\mbox{\tiny struct}}$ by $\frac{1}{s_\alpha}$ times the coefficient for a structure where $q$ is replaced by $q'$, i.e.
\begin{equation}\label{simplified}
\kappa_{\alpha}^{\mbox{\tiny struct}} = \frac{1}{s_{\alpha}}\left<\chi_{\alpha}(g_{p_1})\chi_{\alpha}(g_{p_2})\ldots\chi_{\alpha}(g_p)
\chi^*_{\alpha}(g_{q_1})\chi^*_{\alpha}(g_{q_2})\ldots\chi^*_{\alpha}(g_{q'})\right>.
\end{equation}
If after the reconnection $p$ and $q'$ are identical then we have $g_p=g_{q'}$ allowing us to reduce the number of variables further since $\left<\chi_\alpha(g_p)\chi_\alpha^*(g_p)\right>=1$.

We must now show that performing all other types of reconnection steps leads to the same result, allowing $\kappa_{\alpha}^{\mbox{\tiny struct}} = s_\alpha^{-(L-V)}$ to be infered by recursion.
In one alternative situation $B \cup D$ will contain $l_1^{-1}$ and $l_2$ in two separate orbits indicating that the relevant antiparallel encounter stretches also belong to different orbits. However since the equality $\chi^*_{\alpha}(g^{-1}) = \chi_{\alpha}(g) = \chi^*_{\alpha}(g)$ (for real representations) allows us to invert any of these periodic orbits without fear of altering the group sum we can convert this into a situation with parallel encounter stretches and non-inverted link elements $l_1$ and $l_2$ as treated below.

\subsubsection{Parallel encounter permutations}\label{Parallel_EP}

\begin{figure}[ht]
\centerline{
\includegraphics[width=0.7\paperwidth]{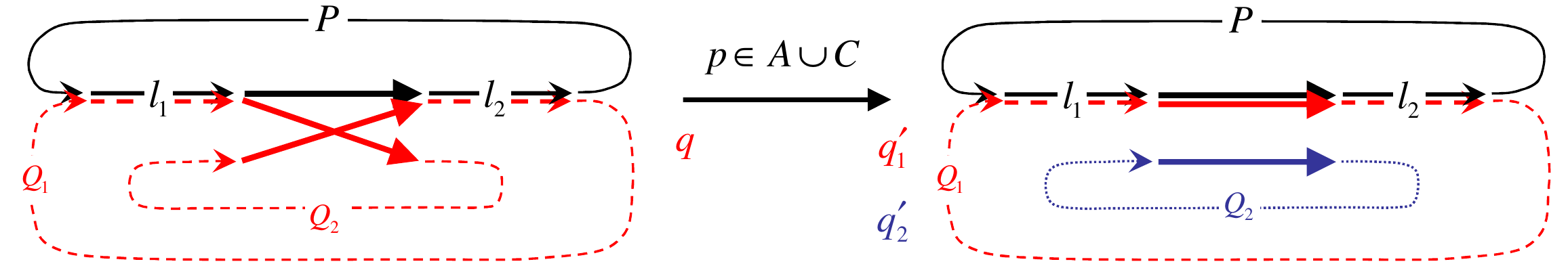}}
\caption{(a) The permutation of parallel encounter stretches within a period orbit $q$ leads to (b) a decomposition of the orbit into two orbits
$q_1'$ and $q_2'$. After the permutation the orbit $q$ has an encounter stretch that is almost identical to a stretch of the orbit $p$.}
\label{Encs02}
\end{figure}
\begin{figure}[ht]
\centerline{
\includegraphics[width=0.7\paperwidth]{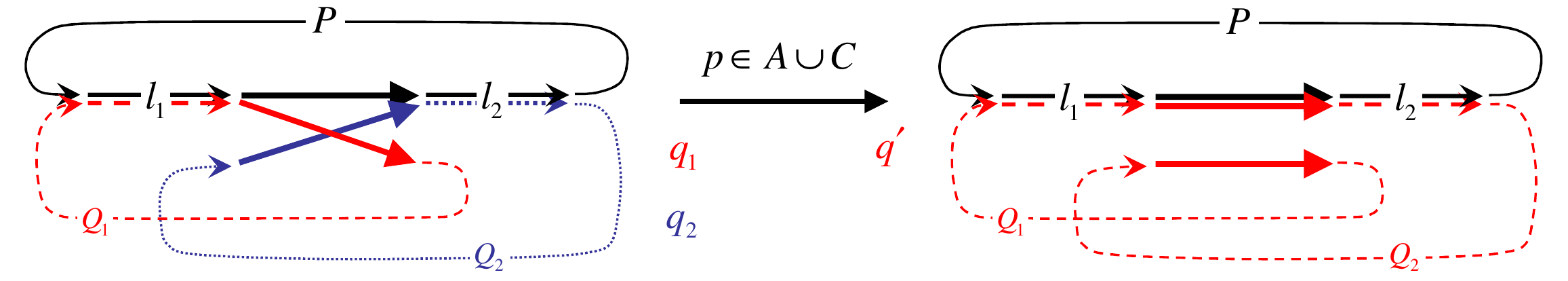}}
\caption{(a) Exchanging two parallel encounters belonging to separate encounter stretches leads to (b) the recombination into one orbit of the combined length.}
\label{Encs03}
\end{figure}

We now deal with the case where $B\cup D$ contains the stretches $l_1$ and $l_2$ without inversion;
the case of $l_1^{-1}$ and $l_2^{-1}$ appearing can be reduced to this situation by reverting all orbits in time.
There are two scenarios: Either $l_1$ and $l_2$ and thus the relevant encounter stretches in $B \cup D$ belong to the same periodic orbit ($q$ say) or they belong to different orbits ($q_1$ and $q_2$ say), illustrated in Figs.~\ref{Encs02} and \ref{Encs03} respectively. In the former case performing the encounter exchange will split the orbit into two separate orbits (denoted by $q'_1$ and $q'_2$), see Fig.~\ref{Encs02}, whereas in the latter case the two orbits will merge into a single orbit (denoted by $q'$), see Fig.~\ref{Encs03}.

We first consider the case where $l_1$ and $l_2$ are contained in the same orbit $q\in B\cup D$. In this case the group elements of $p$ and $q$ can be written in the form
\begin{equation} \label{decomp2}
g_p=Pl_2l_1 \hspace{15pt} \mbox{and} \hspace{15pt} g_q=(Q_2)(l_1 Q_1 l_2).
\end{equation}
As $l_1$ and $l_2$ follow each other in $p$ the orbit $q$ must contain almost parallel encounter stretches following the link with group element $l_1$ and preceding the link with group element $l_2$. The positions of these encounter stretches are indicated by the brackets in the factorisation of $g_q$ in (\ref{decomp2}). Now switching connections between the encounter stretches leads to a decomposition of $q$ into two orbits $q_1'$ and $q_2'$ with group elements
\begin{equation}
g_{q_1'} = l_1 Q_1 l_2 \hspace{15pt} \mbox{and} \hspace{15pt} g_{q_2'}=Q_2 .
\end{equation}
As the links with group elements $l_1$ and $l_2$ follow each other in $q_1'$ just as in $p$ we have thus performed one of the reconnection steps leading from $B\cup D$ to $A\cup C$. As before we now relate the coefficient
$\kappa_{\alpha}^{\mbox{\tiny struct}}$ for our original structure to the one for the structure with $q$ replaced by $q_1'$ and $q_2'$.  The coefficient $\kappa_{\alpha}^{\mbox{\tiny struct}}$  involves the average
\begin{equation}
\frac{1}{|G|^2}\sum_{l_1,l_2\in G} \chi_{\alpha}(Pl_2l_1)\chi^*_{\alpha}(Q_2 l_1 Q_1 l_2).
\end{equation}
We now make the same substitutions  $l_1 \rightarrow x^{-1}l_1$ and $l_2 \rightarrow l_2 x$ as in the previous case to set up an average over $x$ of the form
\begin{equation}
\frac{1}{|G|^2}\sum_{l_1,l_2\in G}\frac{1}{|G|}\sum_{x\in G} \chi_{\alpha}(Pl_2l_1)\chi^*_{\alpha}(Q_2 x^{-1} l_1 Q_1 l_2 x)\,.
\end{equation}
Using the identity $\frac{1}{|G|}\sum_{g \in G}\chi_{\alpha}(agbg^{-1}) = \frac{1}{s_{\alpha}}\chi_{\alpha}(a)\chi_{\alpha}(b)$
(see Eq. (\ref{corr3})) as well as the invariance of the characters under cyclic permutation of the group elements we then obtain
\begin{equation}
\frac{1}{s_{\alpha}}\frac{1}{|G|^2}\sum_{l_1,l_2\in G}\chi_{\alpha}(Pl_2l_1)\chi^*_{\alpha}(l_1 Q_1 l_2)\chi^*_{\alpha}(Q_2)=
\frac{1}{s_{\alpha}}\frac{1}{|G|^2}\sum_{l_1,l_2\in G}\chi_{\alpha}(g_p)\chi^*_{\alpha}(g_{q_1'})\chi^*_{\alpha}(g_{q_2'}).
\end{equation}
If we insert this result into $\kappa_{\alpha}^{\mbox{\tiny struct}}$ we obtain once again a factor of $\frac{1}{s_{\alpha}}$ times the contribution of a structure where one reconnection step has been performed to make $B \cup D$ more similar to $A \cup C$, i.e.
\begin{equation}\label{simplified2}
\kappa_{\alpha}^{\mbox{\tiny struct}} = \frac{1}{s_{\alpha}}\left<\chi_{\alpha}(g_{p_1})\chi_{\alpha}(g_{p_2})\ldots\chi_{\alpha}(g_p)
\chi^*_{\alpha}(g_{q_1})\chi^*_{\alpha}(g_{q_2})\ldots
\chi^*_{\alpha}(g_{q'_1})\chi^*_{\alpha}(g_{q'_2})\right>.
\end{equation}

We should mention that there can be a special case in which the orbit $p$ has only one encounter stretch and one link, in which case the two links $l_1$ and $l_2$ would be identical. In this situation (\ref{decomp2}) is replaced by $g_p=l$ and $g_q=Q l$, and we arrive at the same final result (\ref{simplified2}) if we make the substitution $l \rightarrow x l x^{-1}$.

Now let us assume that the links $l_1$ and $l_2$  belong to two different orbits ($q_1$ and $q_2$) in $B \cup D$ as depicted in Fig.~\ref{Encs03}. In this case the group elements have the form
\begin{equation}
g_p=Pl_2l_1, \hspace{15pt}g_{q_1} = l_1 Q_1 \hspace{15pt} \mbox{and} \hspace{15pt} g_{q_2} = Q_2 l_2.
\end{equation}
By the same arguments as above there are almost parallel encounter stretches following the link element associated to $l_1$ and preceding the link element associated to $l_2$. If we switch connections between these stretches the orbits $q_1$ and $q_2$  merge into a single orbit $q'$ with the group element
\begin{equation}
g_{q'} = (Q_2 l_2)(l_1Q_1).
\end{equation}
Now the coefficient $\kappa_{\alpha}^{\mbox{\tiny struct}}$ involves the average
\begin{equation}
\frac{1}{|G|^2}\sum_{l_1,l_2\in G}\chi_{\alpha}(Pl_2l_1)\chi^*_{\alpha}(Q_1l_1)\chi^*_{\alpha}(l_2Q_2)
=\frac{1}{|G|^3}\sum_{x,l_1,l_2\in G}\chi_{\alpha}(Pl_2l_1)\chi^*_{\alpha}(l_1Q_1x^{-1})\chi^*_{\alpha}(l_2 x Q_2),
\end{equation}
where we have used the invariance of the characters under cyclic permutation of group elements and the same substitutions $l_1 \rightarrow x^{-1}l_1$ and $l_2 \rightarrow l_2 x$ as in the previous two cases. We then use the identity
$\frac{1}{|G|}\sum_{g \in G}\chi_{\alpha}(ag)\chi_{\alpha}(bg^{-1}) = \frac{1}{s_{\alpha}}\chi_{\alpha}(ab)$
(see Eq. \ref{corr2}) and sum over $x$ to get
\begin{equation}
\frac{1}{s_{\alpha}}\frac{1}{|G|^2}\sum_{l_1,l_2\in G}\chi_{\alpha}(Pl_2l_1)\chi^*_{\alpha}(l_1Q_1 Q_2 l_2)
= \frac{1}{s_{\alpha}}\frac{1}{|G|^2}\sum_{l_1,l_2\in G} \chi_{\alpha}(g_p)\chi^*_{\alpha}(g_{q'}).
\end{equation}
Thus after exchanging encounter stretches we again obtain the result that $\kappa_{\alpha}^{\mbox{\tiny struct}}$ is given by $\frac{1}{s_\alpha}$ times the contribution of a structure with one reconnection step removed.

We thus see that performing each of the $L-V$  steps needed to transform the set of periodic orbits in $B \cup D$ into $A \cup C$  produces a factor of $s_{\alpha}^{-1}$. When one of the orbits in $B \cup D$ has been turned into one already appearing in $A \cup C$ this orbit can be dropped from the sum in $\kappa_{\alpha}^{\mbox{\tiny struct}}$
due to $\frac{1}{|G|}\sum_{g\in G}\chi_\alpha(g)\chi_\alpha^*(g)=1$. After all $L-V$ reconnections have been performed no more orbits are left and the value $\kappa_{\alpha}^{\mbox{\tiny struct}}$ for this trivial structure is just 1.
Altogether this implies the desired result
\begin{equation}\label{Final_coeff}
\kappa_{\alpha}^{\mbox{\tiny struct}} = s_{\alpha}^{-(L-V)}
\end{equation}
so Eq. (\ref{sumkappa}) simply becomes
\begin{equation}\label{sumkappafinal}
Z^{(1)}_{\alpha,\mbox{\scriptsize off}}=\sum_{\rm struct}Z^{(1)}_{\mbox{\scriptsize off,\,struct}}\;.
\end{equation}
This sum was already shown to yield GOE behavior in \cite{Mul09}. Hence the correlation function $C_\alpha(\epsilon)$ for subspectra associated to real representations in time reversal invariant systems is
faithful to the GOE prediction, obtained from Eq.~(\ref{GOE_Expansion}) after dropping the restriction
to the real part.

\subsection{Complex Irreducible Representations}
If the representation is complex then the difference between parallel and anti-parallel exchanges becomes greatly significant. For the parallel exchanges in section \ref{Parallel_EP} nothing changes in comparison to the real case, since the group relations (\ref{corr2}) and (\ref{corr3}) hold regardless of the type of representation. However, for the antiparallel exchanges in section \ref{AntiParallel_EP} the group relation (\ref{corr1}) implies a zero contribution when the representation is complex.
The underlying reason for this rests on the inequivalence between a complex representation and its complex conjugated counterpart: An antiparallel exchange implies the existence of a link, say $l_a$, which is traversed in opposite directions in $A\cup C$ and $B\cup D$, corresponding to a time-reversal. If we express the characters as traces of the matrices $M^{(\alpha)}(g)$ and use the representation property $M^{(\alpha)}(g_2 g_1) = 
M^{(\alpha)}(g_2) \, M^{(\alpha)}(g_1)$ then it becomes clear that the average over this one link involves a calculation of the type
\begin{equation} \label{cancel}
\frac{1}{|G|} \sum_{l_a} M_{ij}^{(\alpha)}(l_a) \, M_{kl}^{(\alpha)}(l_a^{-1})^* =
\frac{1}{|G|} \sum_{l_a} M_{ij}^{(\alpha)}(l_a) \, M_{lk}^{(\alpha)}(l_a) =
\frac{1}{|G|} \sum_{l_a} M_{ij}^{(\alpha)}(l_a) \, M_{lk}^{(\beta)}(l_a)^* = 0 \, ,
\end{equation}
where we used in the first step the unitarity of the representation, in the second step $\beta$ denotes the complex conjugate representation
of $\alpha$, and in the third step we used the group orthogonality relation (\ref{grouportho}). The contributions of antiparallel 
exchanges that in the real case could be converted to parallel exchanges by reverting orbits in time vanish for the same reason. Hence only those structures contribute where the periodic orbits in $B\cup D$ follow all links and encounters in the same direction as the periodic orbits in $A\cup C$.
Thus we obtain a sum of the form (\ref{sumkappafinal}) but restricted to structures which do not require time-reversal invariance. This sum was already shown in \cite{Mul09} to yield the GUE behavior.

It is interesting that the structures involving time reversal are dropped not because they are non-existent but because over many pseudo-orbits their average contribution is zero.

\subsection{Non Time-Reversal Invariant Systems} \label{nontime}

For systems without time reversal invariance the structures relying on time reversal are obviously excluded, this time due to the dynamics of the system. We obtain again the result (\ref{sumkappa}) for the off-diagonal contributions, but the sum is now restricted to structures that do not require time-reversal invariance. The calculations in  section \ref{Parallel_EP} on parallel encounter permutations carry over unchanged to show that 
$\kappa_\alpha^{\rm struct}=s^{V-L}_{\alpha}$. As mentioned before, these calculations do not depend on the type of the representation, and in this way one obtains GUE behaviour for both real and complex representations.

\section{Full Correlation Function}\label{Full_corr_function}

To describe the correlations in the full quantum system we need to understand not only correlations inside each subspectrum $\alpha$ but also correlations between different subspectra $\alpha$ and $\beta$. These cross-correlations were also considered in \cite{Bra11} where the leading terms in $\frac{1}{\epsilon}$ as well as the first subleading non-oscillatory term were determined.

When defining the cross-correlation function between the subspectra $\alpha$ and $\beta$ one has to take into account that the
average level densities for the two subspectra can be different, and one could scale the energy according to the average
level densities for either of the subspectra or for the full system. The corresponding dimensionless parameters $\epsilon$,
$\epsilon^{(\alpha)}$, and $\epsilon^{(\beta)}$ are related by $\frac{\epsilon}{\bar\rho}
=\frac{\epsilon^{(\alpha)}}{\bar\rho_\alpha}=\frac{\epsilon^{(\beta)}}{\bar\rho_\beta}$.
For the moment we will use $\epsilon^{(\alpha)}$ and $\epsilon^{(\beta)}$ in parallel and define the cross-correlation function by
\begin{equation} \label{crosscorre}
C_{\alpha\beta}(\epsilon^{(\alpha)},\epsilon^{(\beta)})=\frac{1}{2\pi^2\bar\rho_\alpha\bar\rho_\beta}\left\langle\;
\frac{1}{s_\alpha}{\rm tr}\,P_\alpha\frac{1}{E+\frac{\epsilon^{(\alpha)}}{2\pi\bar\rho_\alpha}-H}\;\;
\frac{1}{s_\beta}{\rm tr}\,P_\alpha\frac{1}{E-\frac{\epsilon^{(\beta)}}{2\pi\bar\rho_\beta}-H}\;
\right\rangle-\frac{1}{2}\;.
\end{equation}
We consider first the case when $\alpha$ and $\beta$ are complex conjugate representations.

\subsection{Scenario 1: \texorpdfstring{$\alpha$ and $\beta$}{} mutually complex conjugate}\label{Scenario_1}

The subspectra associated to two mutually complex conjugate representations are identical in time-reversal invariant systems,
as was discussed in section \ref{Symmetry_in_QM}. As a consequence, their cross-correlation function
(\ref{crosscorre}), with $\epsilon = \epsilon^{(\alpha)} = \epsilon^{(\beta)}$,  must coincide with the correlation function
$C_\alpha(\epsilon)$ for complex representations, i.e., with the GUE result (\ref{GUE_Expansion}).
This can also be seen from semiclassics if we use that two mutually complex conjugate representations $\alpha,\beta$ satisfy $\chi_\beta(g)=\chi_\alpha^*(g)=\chi_\alpha(g^{-1})$. Hence the character associated to an orbit in the representation $\beta$ coincides with the character associated to the time-reversed orbit in $\alpha$. This means that whenever the representation $\beta$ is considered we have to work with time-reversed
orbits instead. As a consequence the result from the diagonal approximation (\ref{diagaverage}) is the same as before because it is unchanged
if partner orbits are replaced by their time-reversed versions. Concerning the off-diagonal contributions, the relevant structures
coincide with the structures used before apart from time reversal of all orbits associated to $\beta$.
These structures are in a one-to-one relation to those considered previously and give the same contributions.

For systems without time-reversal invariance  mutually complex eigenfunctions are not required to possess the same energy and so the two-fold degeneracy occurring between mutually-complex representations disappears. From a semiclassical perspective, this result arises because there
are no time-reversed orbits in our system, and the contributions from parallel encounters vanish because of similar arguments as in (\ref{cancel}).

\subsection{Scenario 2: \texorpdfstring{$\alpha$ and $\beta$}{} not mutually complex conjugate}\label{Scenario_2}

We now consider $\alpha$ and $\beta$ that are neither identical nor mutually complex conjugate. We will show that the corresponding subspectra are uncorrelated in the semiclassical limit. However this does not rule out correlations outside the semiclassical limit as observed in \cite{Dit97}. The generating function that corresponds to the cross-correlation function (\ref{crosscorre}) is defined as
\begin{equation}\label{cross_gen_func}
 Z_{\alpha\beta}\left(\epsilon^{(\alpha)}_A,\epsilon^{(\beta)}_B,\epsilon^{(\alpha)}_C,\epsilon^{(\beta)}_D\right) = \left<
\frac{\Delta_{\alpha}\left(E + \epsilon^{(\alpha)}_C/2\pi\bar{\rho}_{\alpha}\right)
\Delta_{\beta}\left(E - \epsilon^{(\beta)}_D/2\pi\bar{\rho}_{\beta}\right)}
{\Delta_{\alpha}\left(E + \epsilon^{(\alpha)}_A/2\pi\bar{\rho}_{\alpha}\right)
\Delta_{\beta}\left(E - \epsilon^{(\beta)}_B/2\pi\bar{\rho}_{\beta}\right)}\right>
\end{equation}
and gives the cross  correlation function  through
\begin{equation}\label{Cross_cf_access}
C_{\alpha\beta}(\epsilon^{(\alpha)},\epsilon^{(\beta)}) =
-\left.2\frac{\partial^2 Z_{\alpha\beta}}{\partial\epsilon^{(\alpha)}_A\partial\epsilon^{(\beta)}_B}\right|_{(\|)} - \frac{1}{2}.
\end{equation}
where $(\|)$ implies $\epsilon^{(\alpha)}_A=\epsilon^{(\alpha)}_C=\epsilon^{(\alpha)}$,
$\epsilon^{(\beta)}_B=\epsilon^{(\beta)}_D=\epsilon^{(\beta)}$.
The part of the generating function responsible for non-oscillatory contributions is given by
\begin{eqnarray}\label{Full_Gen_func02}
Z^{(1)}_{\alpha\beta} & = & e^{i(\epsilon^{(\alpha)}_A + \epsilon^{(\beta)}_B - \epsilon^{(\alpha)}_C  - \epsilon^{(\beta)}_D)/2}
\left<\sum_{A,B,C,D}G^{(\alpha)}_AG^{(\beta)*}_BG^{(\alpha)}_CG^{(\beta)*}_D(-1)^{n_C+n_D}\right.\nonumber \\
& \times & \left.e^{i(S_A - S_B + S_C - S_D)/\hbar}e^{i(T_A\epsilon^{(\alpha)}_A + T_C\epsilon^{(\alpha)}_C)/T^{(\alpha)}_H}e^{i(T_B\epsilon^{(\beta)}_B  + T_D\epsilon^{(\beta)}_D)/T^{(\beta)}_H}\right>,
\end{eqnarray}
generalizing Eq. (\ref{SC_Generating_Function}). Again the action difference becomes small if the orbits in $B\cup D$ coincide with those in $A\cup C$ apart from their connections inside encounters.
However the orbits in $A\cup C$ contribute with their character in the representation $\alpha$ whereas
the orbits in $B\cup D$ contribute with the complex conjugate of their character in the representation $\beta$.
Hence all contributions vanish due to $\frac{1}{|G|}\sum_{g\in G}\chi_\alpha(g)\chi_\beta^*(g)=0$.
In particular this relation comes into play in Eq. (\ref{diag_coeff}) in the diagonal approximation
and when dealing with orbits that have become identical after the recursion steps in our calculation of
$\kappa_{\alpha}^{\mbox{\tiny struct}}$. Thus  the only contribution to (\ref{Full_Gen_func02}) arises from the empty set, where $n_A = n_B = n_C = n_D = 0$. This summand trivially gives 1 and together with the oscillatory prefactor we have
\begin{equation}
Z^{(1)}_{\alpha\beta} = e^{i(\epsilon^{(\alpha)}_A + \epsilon^{(\beta)}_B - \epsilon^{(\alpha)}_C  - \epsilon^{(\beta)}_D)/2}.
\end{equation}

The second part $Z_{\alpha\beta}^{(2)}$ of the generating function has to be considered separately, since the inclusion of different representations mean it can no longer be obtained directly from $Z_{\alpha\beta}^{(1)}$ through Eq. (\ref{Gen_func_relation}). Instead the complex conjugation of both spectral determinants in the numerator, following Riemann-Siegel resummation, yields
\begin{eqnarray}\label{Full_Gen_func03}
Z^{(2)}_{\alpha\beta} & = & e^{i(\epsilon^{(\alpha)}_A + \epsilon^{(\beta)}_B + \epsilon^{(\alpha)*}_C + \epsilon^{(\beta)*}_D)/2}
\left<\sum_{A,B,C,D}G^{(\alpha)}_AG^{(\beta)*}_BG^{(\alpha)*}_CG^{(\beta)}_D(-1)^{n_C+n_D}\right.\nonumber \\
& \times & \left.e^{i(S_A - S_B - S_C + S_D)/\hbar}e^{i(T_A\epsilon^{(\alpha)}_A - T_C\epsilon^{(\alpha)*}_C)/T^{(\alpha)}_H}e^{i(T_B\epsilon^{(\beta)}_B - T_D\epsilon^{(\beta)*}_D)/T^{(\beta)}_H}\right>.
\end{eqnarray}
The action difference now becomes small if the orbits in $B\cup C$ are related to those in $A\cup D$. However if $\alpha \neq \beta$ and there is (part of) an orbit in $A$ which is correlated to (part of) an orbit in $B$ then the group orthogonality relation for different representations will instill a zero contribution. The same process occurs for those (parts of) orbits in $C$ which are correlated to (parts of) orbits in $D$. The generating function $Z^{(2)}_{\alpha\beta}$ may therefore be split into two uncorrelated factors confined by the two irreducible representations
\begin{eqnarray}
Z^{(2)}_{\alpha\beta} & = & e^{i(\epsilon^{(\alpha)}_A+ \epsilon^{(\alpha)*}_C)/2}
\left<\sum_{A,C}G^{(\alpha)}_AG^{(\alpha)*}_C(-1)^{n_C}e^{i(S_A - S_C)/\hbar}e^{i(T_A\epsilon^{(\alpha)}_A - T_C\epsilon^{(\alpha)*}_C)/T^{(\alpha)}_H}\right> \nonumber \\
& \times &  e^{i(\epsilon^{(\beta)}_B+ \epsilon^{(\beta)*}_D)/2} \left<\sum_{B,D}G^{(\beta)*}_BG^{(\beta)}_D(-1)^{n_D}e^{-i(S_B - S_D)/\hbar}e^{i(T_B\epsilon^{(\beta)}_B - T_D\epsilon^{(\beta)*}_D)/T^{(\beta)}_H}\right>.
\end{eqnarray}
This can once again be separated into the form $Z_{\mbox{\tiny diag}}^{(2)}(1 + Z_{\mbox{\tiny off}}^{(2)})$ where the diagonal part is given by
\begin{eqnarray}\label{crossdiag}
Z_{\alpha\beta,\mbox{\tiny diag}}^{(2)} & = & e^{i(\epsilon^{(\alpha)}_A+ \epsilon^{(\alpha)*}_C)/2} \exp\left(-\kappa^{\mbox{\tiny diag}}_{\alpha}\sum_a|F_a|^2e^{i T_a(\epsilon^{(\alpha)}_A-\epsilon^{(\alpha)*}_C)/T^{(\alpha)}_H}\right) \nonumber \\
& \times & e^{i(\epsilon^{(\beta)}_B+ \epsilon^{(\beta)*}_D)/2}\exp\left(-\kappa^{\mbox{\tiny diag}}_{\beta}\sum_a|F_a|^2e^{i T_a(\epsilon^{(\beta)}_B-\epsilon^{(\beta)*}_D)/T^{(\beta)}_H}\right)\,.
\end{eqnarray}
If we let $x=i(\epsilon^{(\alpha)}_A-\epsilon^{(\alpha)*}_C)$, then ${\rm Re} \, x > 0$, and after invoking Hannay and Ozorio de Almeida's sum rule \cite{Han84}  the sum over $a$ in the first line of (\ref{crossdiag}) turns into
\begin{equation}
\sum_a |F_a|^2e^{ixT/T_H^{(\alpha)}}\sim\int_{T_0}^\infty\frac{dT}{T}e^{ixT/T_H^{(\alpha)}}
=\int_{T_0/T_H^{(\alpha)}}^\infty\frac{d\tau}{\tau}e^{ix\tau}
\end{equation}
and hence diverges in the semiclassical limit due to $T_H^{(\alpha)}\to\infty$. The same applies for the sum in the second line and altogether we have
$Z_{\alpha\beta,\mbox{\tiny diag}}^{(2)}\to 0$.
The off-diagonal part is given by
\begin{eqnarray}\label{full_offdiag}
Z_{\alpha\beta,\mbox{\tiny off}}^{(2)} & = &
\sum_{\mbox{\tiny struct}} \frac{(-1)^{n_C}s_{\alpha}^{L-V}\kappa_{\alpha}^{\mbox{\tiny struct}}}{2^VV!(-1)^L}
\frac{1}{(i \epsilon^{(\alpha)}_A - i \epsilon^{(\alpha)*}_C)^{L-V}} \nonumber \\
& \times & \sum_{\mbox{\tiny struct}} \frac{(-1)^{n_D}s_{\beta}^{L-V}\kappa_{\beta}^{\mbox{\tiny struct}}}{2^VV!(-1)^L}
\frac{1}{(i \epsilon^{(\beta)}_B - i \epsilon^{(\beta)*}_D)^{L-V}}.
\end{eqnarray}
which remains finite even when $\epsilon^{(\alpha)}_A = \epsilon^{(\alpha)}_C$ or $\epsilon^{(\beta)}_B = \epsilon^{(\beta)}_D$ in the denominator.
Thus $Z_{\alpha\beta}^{(2)}$ can indeed be dropped leaving only
\begin{equation}
Z_{\alpha\beta} = Z_{\alpha\beta}^{(1)} = e^{i(\epsilon^{(\alpha)}_A + \epsilon^{(\beta)}_B - \epsilon^{(\alpha)}_C - \epsilon^{(\beta)}_D)/2}
\end{equation}
and thus combined with relation (\ref{Cross_cf_access})
\begin{equation}
C_{\alpha\beta}(\epsilon^{(\alpha)},\epsilon^{(\beta)}) =
-\left.2\frac{\partial^2}{\partial\epsilon^{(\alpha)}_A\partial\epsilon^{(\beta)}_B}\left[
e^{i(\epsilon^{(\alpha)}_A + \epsilon^{(\beta)}_B - \epsilon^{(\alpha)}_C - \epsilon^{(\beta)}_D)/2}\right]
\right|_{(\|)} - \frac{1}{2} = 0.
\end{equation}
Hence in the semiclassical limit there are no correlations between subspectra associated to different
representations unless these representations are mutually complex conjugate.

\subsection{Final result}\label{final_result}

The correlation function of the full system is given by a linear combination of all correlation functions
inside and between subspectra. Taking into account the scaling factors from
$\rho=\sum_\alpha s_\alpha\rho_\alpha$ and $\epsilon^{(\alpha)}=\frac{\bar\rho_\alpha}{\bar\rho}\epsilon=\frac{s_\alpha}{|G|}\epsilon$ 
we obtain
\begin{equation}\label{full_cf_sum}
C(\epsilon) = \sum_{\alpha,\beta}\frac{s_\alpha\bar{\rho}_{\alpha}s_\beta\bar{\rho}_{\beta}}{\bar{\rho}^2}C_{\alpha\beta}(\epsilon^{(\alpha)},\epsilon^{(\beta)})
=\sum_{\alpha,\beta}\frac{s_\alpha^2s_\beta^2}{|G|^2}C_{\alpha\beta}\left(\frac{s_\alpha}{|G|}\epsilon,\frac{s_\beta}{|G|}\epsilon\right)
\end{equation}
For time-reversal invariant systems we have shown that
\begin{equation}
C_{\alpha\beta}(\epsilon^{(\alpha)},\epsilon^{(\beta)})=\begin{cases}
C_{\rm GOE}(\epsilon^{(\alpha)})&\text{if $\alpha=\beta$ real}\\
C_{\rm GUE}(\epsilon^{(\alpha)})&\text{if $\alpha=\beta$ complex or $\alpha$, $\beta$ mutually complex}\\
0&\text{otherwise}
\end{cases}
\end{equation}
yielding
\begin{equation}\label{final_result_2}
C(\epsilon) =
\frac{1}{|G|^2}\left[\sum_{\alpha\;{\rm real}} s^4_{\alpha}C_{\mbox{\tiny GOE}}\left(\frac{s_\alpha}{|G|}\epsilon\right)
+ \sum_{\alpha\;{\rm complex}} 2 s^4_{\alpha}C_{\mbox{\tiny GUE}}\left(\frac{s_\alpha}{|G|}\epsilon\right)\right].
\end{equation}
For systems without time-reversal invariance
we have shown that
\begin{equation}
C_{\alpha\beta}(\epsilon^{(\alpha)},\epsilon^{(\beta)})=\begin{cases}
C_{\rm GUE}(\epsilon^{(\alpha)})&\text{if $\alpha=\beta$}\\
0&\text{otherwise}
\end{cases}
\end{equation}
and thus
\begin{equation}\label{final_result_NTR}
C(\epsilon) =
\frac{1}{|G|^2}\sum_{\alpha} s^4_{\alpha}C_{\mbox{\tiny GUE}}\left(\frac{s_\alpha}{|G|}\epsilon\right).
\end{equation}

This is our final result: It shows that a time-reversal invariant quantum chaotic system has a correlation function in the semiclassical limit that is comprised of sums of GOE and GUE correlation functions corresponding to real and complex subspaces and weighted by appropriate degeneracy factors (provided that the fundamental domain does not possess any hidden symmetries leading to deviations from RMT as in arithmetical billiards \cite{Bog96b}). In contrast, a non-time reversal invariant system has no additional degeneracy factor due to complex conjugate representations and every subspace has a GUE correlation function.

\section{Conclusions}

\label{Conclusions}

For chaotic systems with a discrete spatial symmetry we have shown that periodic-orbit theory can account for correlations of
levels inside each of the subspectra that are associated with the different irreducible representations $\alpha$ of the symmetry group 
as well as for cross-correlations between these subspectra.
This was achieved by considering the semiclassical realisation of the symmetry-reduced spectral determinant $\Delta_{\alpha}(E)$
and implementing the Riemann-Siegel lookalike formula, from which the exact asymptotic RMT expansions for the
complex correlation functions  were successfully reproduced. The results are summarised in Table~\ref{tab:results}.
For real representations in time-reversal invariant systems we obtained GOE behaviour and otherwise GUE behaviour.

\begin{table*}[th]
\begin{center}
\begin{tabular}{|c|c|c|} \hline \str
                             & no $T$ inv. & $T^2=1$ \\[0.3ex] \hline \str 
$\alpha$ complex & GUE    & GUE       \\[0.3ex] \hline \str 
$\alpha$ real        & GUE    & GOE       \\[0.3ex] \hline
\end{tabular}
\end{center}
\caption{\label{tab:results} Table showing how the random matrix ensembles for subspectra depend on representation
type as well as time reversal properties of the system.}
\end{table*}

Throughout this paper we have neglected the third type of irreducible representations, known as pseudo-real (with complex 
representation matrix $M^{(\alpha)}$ but real trace $\chi_\alpha$). These arise for symmetries which are more complex than the
standard crystallographic point groups, with the simplest being the quaternion group $Q8$. However, from a semiclassical
analysis this case is interesting; showing, contrary to earlier predictions \cite{Kea97}, to possess a GSE distribution rather
than GOE \cite{Joy11}. We have also neglected the case of time-reversal operators squaring to minus unity $T^2=-1$ as is
the case in systems with half-integer spin. For these systems we expect GUE for complex representations, GSE for real
representations, and GOE for pseudo-real representations. Including these cases would extend Table~\ref{tab:results}
to a $3 \times 3$ table.

The techniques in this paper could also be used to analyse the effects of false time reversal symmetry \cite{Ber86b}, continuous symmetries \cite{Cre90}, symmetry breaking \cite{Cre96,Whi09a,Whi09b}, and arithmetical symmetries such as the eigenvalues of the Laplace-Beltrami operator on the domain of the modular group \cite{Bog96b}.
\vspace{10pt}

\noindent \textbf{Acknowledgements:} The authors are grateful to B. Gutkin for relevant discussion and correspondence. C. Joyner would also like to thank J. Robbins for useful advice and comments regarding this work.

\end{document}